\documentclass[aps,prb,10pt,twocolumn,showpacs,superscriptaddress,longbibliography]{revtex4-2}
\usepackage{amsmath} 
\usepackage{amssymb}
\usepackage{physics}
\usepackage[normalem]{ulem}
\usepackage{enumitem}
\usepackage{placeins}
\usepackage{graphicx,epstopdf}
\usepackage[dvipsnames]{xcolor}
\usepackage[T1]{fontenc}
\usepackage[utf8]{inputenc}
\newcommand{\be}{\begin{equation}}
\newcommand{\ee}{\end{equation}}
\newcommand{\bea}{\begin{eqnarray}}
\newcommand{\eea}{\end{eqnarray}}
\newcommand{\bse}{\begin{subequations}}
\newcommand{\ese}{\end{subequations}}

\usepackage{color}
\usepackage[colorlinks,bookmarks=false,citecolor=darkblue,linkcolor=red,urlcolor=blue]{hyperref}

\definecolor{darkred}{rgb}{0.7,0.0,0.0}

\definecolor{darkblue}{rgb}{0,0.02,0.45}

\definecolor{darkgreen}{rgb}{0.02,0.45,0.0}

\definecolor{violet}{rgb}{0.8,0.2,0.6}

\begin{document}

\title{Interplay of disorder and spin liquid physics in the triangular-lattice antiferromagnet 
 YbZn$_2$GaO$_5$}

\author{Hafsa Zeroual}\altaffiliation{These authors contributed equally to this work.}
\affiliation{Université Paris-Saclay, CNRS, Laboratoire de Physique des Solides, 91405, Orsay, France}
\author{Shams Sohel Islam}\altaffiliation{These authors contributed equally to this work.}
\affiliation{PSI Center for Neutron and Muon Sciences CNM, 5232 Villigen PSI, Switzerland}
\author{Toni Shiroka}
\affiliation{PSI Center for Neutron and Muon Sciences CNM, 5232 Villigen PSI, Switzerland}
\author{Thomas James Hicken}
\affiliation{PSI Center for Neutron and Muon Sciences CNM, 5232 Villigen PSI, Switzerland}
\author{Rabindranath Bag}
\affiliation{Department of Physics, Duke University, Durham, NC, USA}
\author{Sara Haravifard}
\affiliation{Department of Physics, Duke University, Durham, NC, USA}
\author{Hubertus Luetkens}
\affiliation{PSI Center for Neutron and Muon Sciences CNM, 5232 Villigen PSI, Switzerland}
\author{Philippe Mendels}
\affiliation{Université Paris-Saclay, CNRS, Laboratoire de Physique des Solides, 91405, Orsay, France}
\author{Edwin Kermarrec}
\affiliation{Université Paris-Saclay, CNRS, Laboratoire de Physique des Solides, 91405, Orsay, France}
\author{Fabrice Bert}\email[Contact author: ]{fabrice.bert@universite-paris-saclay.fr}
\affiliation{Université Paris-Saclay, CNRS, Laboratoire de Physique des Solides, 91405, Orsay, France}
\date{\today}

\begin{abstract}
YbZn$_2$GaO$_5$ has recently emerged as a structurally ideal triangular-lattice antiferromagnet and a promising candidate for hosting a U(1) Dirac quantum spin-liquid (QSL) ground state. However, signatures of broadened crystal electric field (CEF) excitations have raised questions about potential site disorder and its impact on the intrinsic spin dynamics. In this study, we employ $^{71}$Ga nuclear magnetic resonance (NMR) and muon spin relaxation ($\mu$SR) to microscopically probe both the structural and magnetic properties of YbZn$_2$GaO$_5$. NMR spectra reveal two inequivalent $^{71}$Ga sites and a broad distribution of quadrupolar frequencies, providing clear evidence of Ga$^{3+}$/Zn$^{2+}$ site mixing and local CEF randomness. The NMR spin-lattice relaxation rate exhibits, down to 1.3~K, a power-law temperature dependence, consistent with persisting spin dynamics below the exchange energy scale in a frustrated magnet. 
Complementary $\mu$SR measurements further show that the spin dynamics persist down to 30\,mK, with no signs of spin freezing. Together, these results establish that YbZn$_2$GaO$_5$ presents substantial structural disorder, while the persistence of dynamic spin fluctuations down to the lowest temperatures may indicate the survival of QSL-like excitations despite such disorder.
\end{abstract}

\maketitle

\section{Introduction}
The quantum spin-liquid (QSL) state is an exotic, fluid-like phase of matter in which localized magnetic moments evade conventional symmetry breaking even at absolute zero temperature. Rather than forming a static order, the electronic spins remain highly entangled and fluctuating, giving rise to fractionalized excitations, such as spinons, and often feature emergent gauge structures and nontrivial topology~\cite{Broholm2020}. Anderson first proposed the QSL concept in 1973 in the context of the geometrically frustrated $S = 1/2$ antiferromagnetic triangular-lattice Heisenberg model (THM), envisioning a resonating valence bond (RVB) ground state—a quantum superposition of singlets on nearest-neighbor bonds~\cite{Anderson1973}. Although later numerical studies showed that an isotropic THM achieves a 120$^\circ$ Néel order, its strongly suppressed sublattice magnetization (here, the order parameter) compared to the classical limit indicates proximity to a quantum critical point between ordered and disordered phases~\cite{Singh1992, Capriotti1999}. This has continued to motivate experimental efforts to realize QSLs by tuning competing many-body interactions in real materials.

Among the promising platforms, rare-earth triangular-lattice antiferromagnets (TLAFs) have attracted significant interest due to the combined effects of spin-orbit coupling (SOC), crystal electric-field (CEF) interactions, and geometric frustration. Together, these factors can stabilize effective $J_{\text{eff}} = 1/2$  Kramers doublets with anisotropic exchange, providing a basis for quantum spin liquid (QSL) behavior~\cite{Zhu2018, Maksimov2019}. A prominent example is YbMgGaO$_4$, which exhibits no clear magnetic ordering down to 50\,mK and a broad, continuum-like inelastic neutron scattering (INS) spectrum—both hallmarks of a QSL~\cite{Li167203, Paddison117}. However, its interpretation is complicated by chemical disorder arising from random Ga$^{3+}$/Mg$^{2+}$ site occupancy. Such disorder has been proposed to affect the local crystal electric field (CEF) environment of Yb$^{3+}$, leading to a broad distribution of $g$-factors and exchange couplings~\cite{Li107202}. Recent theoretical studies further suggest that this disorder may induce quantum spin-liquid (QSL)-like mimicry~\cite{Zhu157201} or stabilize a valence-bond glass phase~\cite{Kimchi031208}.
The structurally related YbZnGaO$_4$ exhibits even broader excitations and is proposed to host a spin-glass-like~\cite{Ma087201, Ma224433} ground state. Nonetheless, recent thermal conductivity measurements in YbMgGaO$_4$ reveal a residual linear term at ultra-low temperatures, suggesting the presence of mobile, gapless spin excitations, thus reinforcing the gapless QSL scenario~\cite{Rao2021}. This view is further supported by INS results, showing that low-energy excitations stem from valence bond rearrangements, pointing to the propagation of unpaired spins as the origin of the observed continuum~\cite{Li137201}. These findings highlight the need for cleaner triangular-lattice systems to disentangle intrinsic QSL physics from disorder effects.

A promising recent development in this field is the identification of YbZn$_2$GaO$_5$ as a triangular-lattice compound of Yb$^{3+} (J_{\text{eff}} = 1/2)$ ions, initially proposed to be free of any structural disorder~\cite{Bag266703}. Thermodynamic measurements on single crystals reveal a low-temperature specific heat that depends on $T^2$ and with a field-induced linear term, both features being consistent with the  theoretical expectations for a U(1) Dirac QSL~\cite{Hu207203}. Inelastic neutron scattering~\cite{Bag266703} reveal a broad continuum of excitations centered at the $M$ and $K$ points in momentum space and a suppression of spectral weight near the $\Gamma$ point—ruling out a spinon Fermi surface, but supporting a Dirac QSL description. Later $\mu$SR studies further confirmed this scenario~\cite{wu2025spin}. Yet, despite promising thermodynamic and spectroscopic signatures of a Dirac quantum spin-liquid ground state in YbZn$_2$GaO$_5$, recent reports of broadened crystal electric field (CEF) excitations~\cite{zhao2025quenched} have raised concerns about possible Ga$^{3+}$/Zn$^{2+}$ site disorder—analogous to Yb(Mg/Zn)GaO$_4$. 
Further, the absence of spin-wave measurements at high fields in the fully-polarized phase hinders a precise determination of the degree of disorder.

Given the weak contrast $\approx 25\%$ between
Zn and Ga in neutron diffraction, the refinement of powder data~\cite{Bag266703} does not allow us to reliably determine their site occupancies. Hence, it becomes essential to employ complementary techniques to obtain further insight, such as through a local probe. In this study, conducted on single crystals originating from the same growth batch as Ref.~\onlinecite{Bag266703}, we 
use $^{71}$Ga NMR, which offers a unique site-specific sensitivity to local electric-field gradients and hyperfine fields, making it a powerful tool for uncovering static site disorder and CEF randomness. The NMR spectra give clear evidence of Ga$^{3+}$/Zn$^{2+}$ site mixing, manifested as two distinct peaks, corresponding to two inequivalent $^{71}$Ga crystallographic sites—deviating from expectations for the reported ideal structure~\cite{Bag266703}. Moreover, a broad distribution of quadrupolar frequencies indicates significant local charge disorder. To probe the low-energy spin dynamics, we further utilize both NMR and $\mu$SR. The $^{71}$Ga NMR spin-lattice relaxation rate exhibits a temperature dependence down to 1.3~K characteristic of 
magnetic excitations in a frustrated system. Complementary zero-field and longitudinal-field $\mu$SR measurements confirm the persistence of dynamic spin fluctuations down to 30\,mK. Together, these results demonstrate that YbZn$_2$GaO$_5$ hosts a magnetically disordered, dynamically fluctuating ground state with substantial disorder. Finally, we discuss the implications of this disorder for the nature of the quantum magnetic ground state.

\section{NMR results}
\begin{figure*}
	\includegraphics[width = \linewidth]{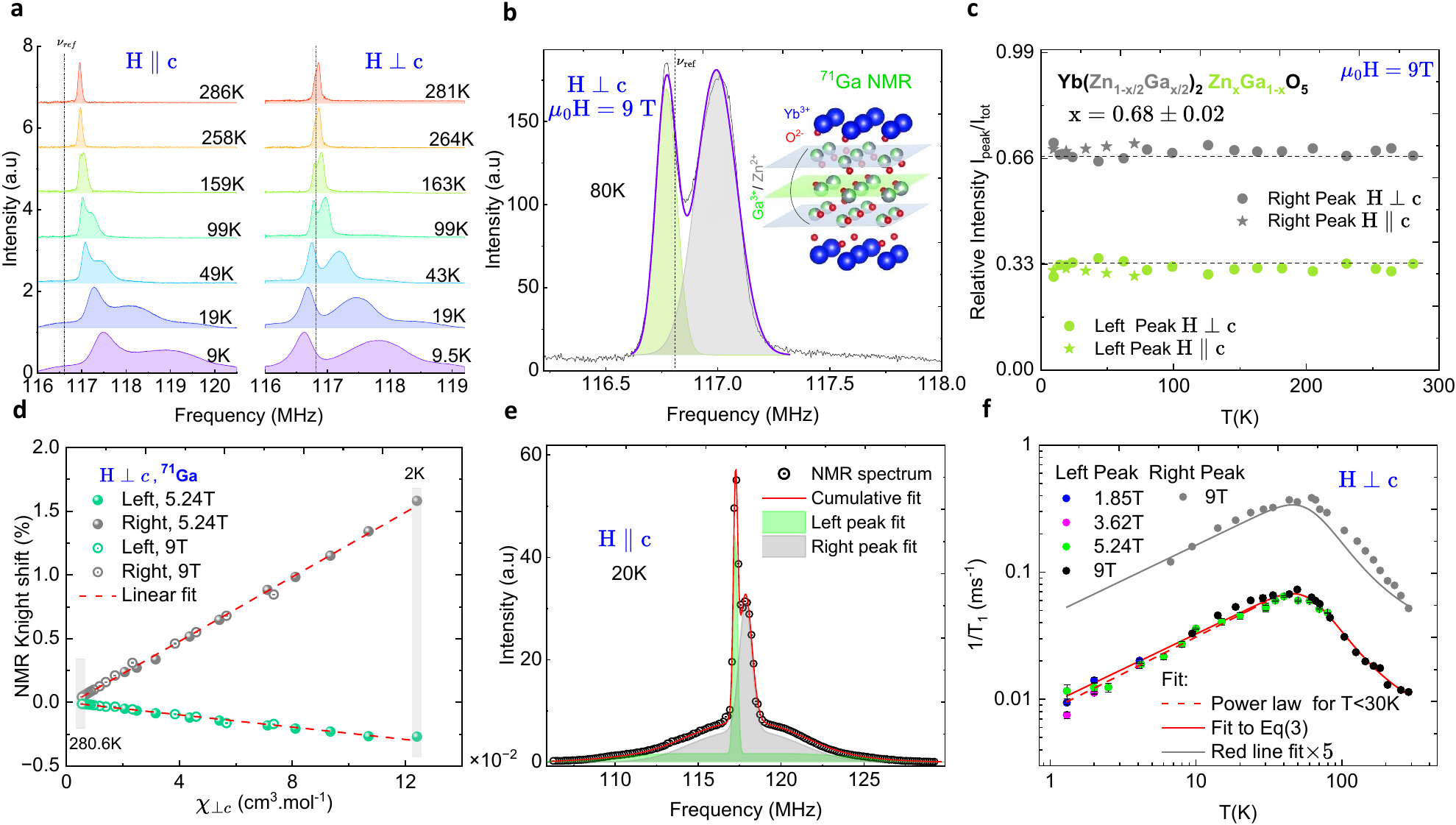}
	\caption{\label{spectral}\textbf{a} Evolution with temperature of the $^{71}$Ga NMR central line normalized to the maximum for $H\parallel c$ and $H \perp c$, measured at 8.996\,T. The dashed vertical line at $\nu_\mathrm{ref} = 116.808$\,MHz indicates the reference frequency.
    \textbf{b} At low temperatures, the $^{71}$Ga NMR central line splits into two and is fitted with two Gaussians (purple curve). Filling colors reflect the two distinct  crystallographic sites. \textbf{c} Relative weights of the two central lines. \textbf{d} NMR line shift of the two sites for $H\perp c$ as a function of the bulk susceptibility measured at 5~T. The temperature range is 2--280\,K, while the applied magnetic field is 8.996 or 5.24\,T. Red lines represent linear fits of $K$ vs $\chi$.  \textbf{e} Complete $^{71}$Ga NMR spectrum showing two central lines and broad satellites. The red line represents the cumulative fit obtained from the sum of three-Gaussian fit applied to each site (filled areas). \textbf{f} Temperature evolution of the spin-lattice relaxation rate $1/T_1$ at different applied fields. The lower curve corresponds to the left-peak data, with the red line showing the fit to Eq.~\ref{crossover}. 
    The red dashed line is a power-law fit in the range 1.3--30\,K. The upper curve shows the right-peak data. The gray line reproduces the left-peak fit scaled by a factor 5.}
\end{figure*}
Here, we report our $^{71}$Ga NMR investigation carried out on a 45-mg single crystal of YbZn$_2$GaO$_5$ with an external magnetic field $B_\mathrm{ext}$ applied parallel and perpendicular to the crystallographic $c$ axis. 
We first discuss the spectra measured at a fixed field of 8.996\,T in the temperature range 9--300\,K, and at 5.24 \,T in the temperature range 1.3--80\,K. 
We performed single-zone acquisitions for the narrow peaks and frequency sweeps for the broad ones, using a $\pi/2-\tau-\pi$ pulse sequence with $\tau=30\mu$s much shorter than the transverse relaxation times. The repetition time was adjusted at each temperature to ensure that the spin lattice relaxation time would not affect the signal intensity. The reference frequency was measured using a Ga(OH)$_3$ solution, yielding $\nu_\mathrm{ref}=116.808$\,MHz for $^{71}$Ga at 8.996\,T.\\
Figure~\ref{spectral}(a) shows the temperature evolution of the central line corresponding to the nuclear $-1/2 \rightarrow +1/2$ transition: at room temperature, we observe an asymmetric line shape, 
whose asymmetry is more pronounced for $B_\mathrm{ext} \perp c$. As we cool down, the single line splits into two distinct peaks that shift differently, depending on the orientation of the external field. This observation shows the existence of two inequivalent $^{71}$Ga crystallographic sites, at variance with what is expected based on the ideal crystal structure reported in Ref.~\onlinecite{Bag266703}. 

To determine the occupancy of each site, the NMR spectra were fitted using two Gaussian curves, as depicted in Fig.~\ref{spectral}(b). In the $B_{\rm ext} \parallel c$ configuration, the weights of the Gaussians were kept free only below $70$K, where the peaks are sufficiently resolved. In the more favorable $B_{\rm ext} \perp c$ configuration, the weights were set free for all temperatures. 
We  checked that two other crystals from the same batch have the same spectra around 80~K and therefore the same intensity ratio for the two lines.
The fits yield a fraction ${I_{\rm left}}/{I_{\rm total}} \approx 1/3$ for the left peak and, thus, ${I_{\rm right}}/{I_{\rm total}} \approx 2/3$ for the right peak [see Fig.~\ref{spectral}(c)]. 

The shift of the NMR 
line is given by $\displaystyle \frac{\nu - \nu_0}{\nu_0} = K_{\rm spin} + K_{\rm chem}$, where $K_{\rm chem}$ is the temperature independent chemical shift and $K_{\rm spin}$ is the shift arising from the polarization of the unpaired electron spins of neighboring Yb$^{3+}$ ions, the latter being proportional to the spin susceptibility. Figure~\ref{spectral}(d) depicts the shifts of both NMR peaks vs bulk susceptibility, for $B_{\rm ext} \perp c$, measured at $\sim$9\,T (from  9.5\,K to 281\,K) and at $\sim$5\,T (from 2\,K to 50\,K). The bulk susceptibility was measured at 5T on the same crystal using a MPMS3 SQUID magnetometer. The NMR lineshifts exhibit a field-independent linear behavior, whose slope corresponds to the hyperfine coupling constant at each $^{71}$Ga site. After correcting for demagnetizing field contribution (see Supplemental Material~\cite{SM}), we obtain $A_{\rm hf}^{\rm L} = - 6$\,mT/$\mu_{\rm B}$ and $A_{\rm hf}^{\rm R} = 78$\,mT/$\mu_{\rm B}$, which indicate that the gallium atoms at the crystallographic site corresponding to the right peak are more strongly coupled to the Yb$^{3+}$ planes than those at the crystallographic site corresponding to the left peak.
   
Furthermore, since $^{71}$Ga possesses a nuclear spin $I = 3/2$, it has a finite quadrupolar moment $Q$. This implies an electrostatic interaction between the nuclear charge distribution and the electric field gradient due to the surrounding electronic charges. The quadrupolar interaction modifies the nuclear Zeeman transitions $-3/2 \rightarrow -1/2$ and $+1/2 \rightarrow +3/2$, leaving the central  $-1/2 \rightarrow +1/2$ transition unchanged (to first order). Thus, the overall $^{71}$Ga NMR spectrum should consist of a central line and two satellites, each shifted by the quadrupolar frequency $\pm \nu_{Q\rm }$ from the central line. 
Our point-charge calculations, based on the crystallographic structure reported in Ref.~\cite{Bag266703}, indicate that the principal axis of the electric field gradient (EFG) tensor is aligned along the crystallographic $c$ axis, with a vanishing asymmetry parameter. This result is consistent with the axial symmetry imposed by the $\overline{6}m2$ local point group symmetry of the Ga crystallographic site (2b). In this case, for $B_{\rm ext}\parallel c$, $\nu_{Q\rm }$ is given by~\cite{slichter2013principles}: 
\begin{equation}
	\nu_Q = \frac{3e^2qQ}{2I(2I-1)h},
\end{equation} 
where $eq$ is a parameter proportional to the largest eigenvalue of the electric field-gradient tensor, with a strength depending on the symmetry of the surrounding charge distribution.
In order to investigate the quadrupolar contribution, we swept the applied field at 20\,K, while maintaining a constant irradiation frequency $\nu_0 = 116.808~\mathrm{MHz}$. For better readability, the spectrum is presented in frequency units using the conversion $\nu = 2\nu_0 -\frac{\gamma}{2\pi}\,B$, where $\gamma= 2\pi \times 12.984~\mathrm{MHz/T}$ is the gyromagnetic ratio of~ $^{71}$Ga and $B$ is the value of the swept field. Figure~\ref{spectral}(e) shows the obtained spectrum, where only a broad quadrupolar background is observed beyond the central transition lines, thus  signaling a large distribution of quadrupolar frequencies. Each site contribution was qualitatively modeled using three Gaussian components: one for the central transition and two for the quadrupolar satellites. The satellites are positioned symmetrically at $\nu_0^{L(R)} \pm \nu_Q^{L(R)}$ with respect to the central lines located at $\nu_0^{L(R)}$. Thus, $\nu_Q^{L(R)}$ corresponds to the average quadrupolar frequency, while the distribution of the quadrupolar frequencies is accounted for by the gaussian full width at half maximum $w_Q^{L(R)}$, which provides a quantitative measure of the spread in $\nu_Q$. The total spectrum is obtained as the sum of the contributions from the two sites. The extracted parameters are $\nu_Q^L = 3.8(5)\,\text{MHz}$ and $w_Q^L = 4.2\,\text{MHz}$ for the left peak, and $\nu_Q^R = 1.7(2)\,\text{MHz}$ and $w_Q^R = 2.7\,\text{MHz}$ for the right peak.
Such a broad distribution of quadrupolar frequencies reflects an \emph{intrinsic random charge distribution} within the single crystal. As in similar compounds~\cite{Li107202}, and in agreement with a recent neutron study~\cite{zhao2025quenched}, our results can be interpreted as Ga$^{3+}$/Zn$^{2+}$ site mixing. The resulting random charge environment not only distributes the quadrupolar frequencies but also leads to a distribution of orientations of the EFG principal axis and asymmetry parameter. Such randomness also induces atomic displacements~\cite{zhao2025quenched} that can further broaden the electric field gradient distribution.\\ 
Therefore, based on our spectral analysis and maintaining the charge balance of the compound, we conclude that the single crystal is best described by the chemical formula: Yb(Zn$_{1-\frac{x}{2}}$Ga$_{\frac{x}{2}}$)$_2($Zn$_{x}$Ga$_{1-x})$O$_5$, with $x = 0.68 \pm \,0.02$, while the ideal structure would correspond to $x=0$.
From our Knight shift results, the right line can be assigned to gallium atoms occupying the (4f) Wyckoff site (Zn site, with local point-group symmetry (3m), gray plane), which is located closer to the Yb$^{3+}$ planes [see Fig.~\ref{spectral}(b)]. In contrast, the left line, which exhibits a weaker hyperfine coupling, can be attributed to gallium atoms at the (2b) Wyckoff site (green plane), situated farther from the Yb$^{3+}$ planes.     

To investigate the magnetic fluctuations, we used an inversion-recovery sequence to measure the $1/T_1$ spin-lattice relaxation rate vs temperature (on both Ga sites) in an applied field $B_\mathrm{ext} \perp c$ [see  Fig.~\ref{spectral}(f) and Supplemental Material~\cite{SM}]. Due to the  inhomogeneous relaxation of NMR lines, for each temperature, the irradiation frequency was set at the peaks position. Since the left peak shows a transverse relaxation time $T_2$ almost three times that of the right peak, we used a longer delay between pulses in the spin-echo sequence in order to reduce the contribution of the right peak. Further, applying the magnetic field in the $B_\mathrm{ext} \perp c$ direction improves the separation of the two central lines, thanks to their opposite shifts.
We fitted the recovery curves of the central lines to the relaxation function \cite{suter1998mixed}:
\begin{equation}
  M(t) = M_\mathrm{sat} \left[1-A\,(0.1e^{-(\frac{t}{T_1})^\beta} + 0.9e^{-(\frac{6t}{T_1} )^\beta }) \right],
\end{equation}
where $\beta \leq 1$ is a phenomenological stretching parameter used to capture the distribution of $T_1$ values, while $A$ is a factor accounting for the imperfect inversion of the spins (ideally, $A \sim 2$).
Below 30\,K, all the recovery curves were fitted using an additional contribution from first order quadrupolar background at the peak position.  
Indeed, such contribution becomes important at low temperatures because the broadening of the central lines increases as the susceptibility, whereas the width of the quadrupolar signal is dominated by disorder, the latter being temperature-independent. We performed $T_1$ measurements on the left peak at 9\,T and 5.24\,T in the temperature ranges 9.5--300\,K and 1.3--80\,K, respectively, and at lower fields (3.62\,T and 1.85\,T) at 1.3, 2, and 4.2\,K. In the temperature range 50--300\,K, we observe an activated regime, 
as reported in \cite{wu2025spin}. For $T \leq 30$\,K,  instead, $1/T_1$ increases as a power law with an exponent $0.58(3)$ [red dashed line in Fig.~\ref{spectral}(f)], 
without any change at the Curie-Weiss temperature $\theta_{\perp} \approx 5$\,K \cite{Bag266703} showing the frustrated nature of the compound. The low-field measurements show no clear field dependence, indicating the absence of saturation effects, consistent with magnetization measurements under applied magnetic field~\cite{Bag266703}. Following Ref.~\cite{wu2025spin}, all our data could be fitted to the empirical function 
\begin{equation}
	\frac{1}{T_1} = \frac{1}{aT^\alpha + b\exp(-\frac{E_A}{k_\mathrm{B}T})},
\label{crossover}
\end{equation}
which 
reflects a crossover between two dominant relaxation regimes: thermally activated at high temperature and power-law at low temperature.
The resulting fit [red line in Fig.~\ref{spectral}(f)] captures quite well the temperature evolution of the relaxation and yields an activation energy $E_A = 19(1)$\,meV. This value is broadly consistent with a former $\mu$SR study performed with a negligible field of 0.01~T~\cite{wu2025spin}, so we can rule out a field induced mechanism, as confirmed by the comparatively small Zeeman energy $\approx 1.6$~meV at 9~T. The high temperature decrease of $1/T_1$, with its characteristic energy scale $E_A$,  likely results from a Raman-like multi-phonon relaxation process involving the first excited CEF level at 38\,meV.
We also obtain the exponent $\alpha = -0.55(3)$ characterizing the low temperature regime. Noticeably the relaxation rate remains algebraic in $T$, even below $\theta_{CW}$, the onset of the correlated regime. 
Our measurements at 9\,T show that the left- and the right peak exhibit the same $1/T_1$ temperature dependence, consistent with the interpretation that gallium atoms in both sites probe the same magnetic fluctuations, differing only in their coupling strength to the Yb$^{3+}$ planes. 

In summary, our NMR spectral analysis clearly shows that there is Ga$^{3+}$/Zn$^{2+}$ site mixing, which leads to a random charge distribution. This is reflected in the wide range of quadrupolar frequencies. Spin-lattice relaxation measurements reveal a temperature dependence down to $ \sim \theta_{\rm CW}/4$ indicating the persistence of low-energy spin dynamics.
Further measurements at much lower temperatures 
and fields could show if other possible low-temperature regimes~\cite{ranjith2019field} emerge, including the Dirac spin liquid one below $\approx$0.1~K~\cite{Bag266703}. They would be interesting to enable a direct comparison with heat capacity~\cite{Bag266703} and the following \textmu SR results in the correlated regime.


     
\section{{\textmu}SR results}
\begin{figure*}
	\includegraphics[width = \linewidth]{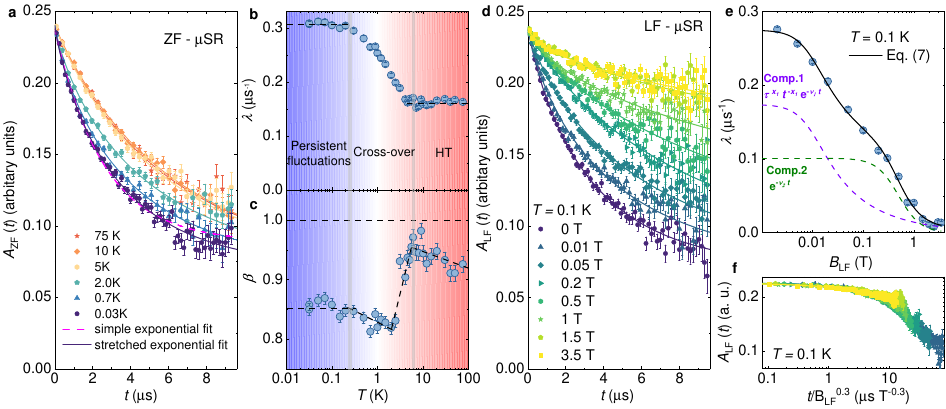}
	\caption{\label{Fig1}\textbf{a} Temperature evolution of the ZF muon spin asymmetry, $A_{\rm ZF}(t)$ for YbZn$_2$GaO$_5$. Dashed and solid lines represent fits using simple- and stretched-exponential functions. \textbf{b} The temperature dependence of the muon-spin relaxation rate suggests three distinct regimes: high temperature (HT), crossover, and persistent fluctuations. \textbf{c} Temperature dependence of the stretch exponent $\beta$. \textbf{d} Muon-spin asymmetry curves $A_{\rm LF}(t)$ for different longitudinal fields at a fixed temperature, $T\simeq0.1$\,K. \textbf{e} Longitudinal field (LF) dependence of the muon-spin relaxation rate $\lambda$. The solid line represents a two-component fit [power-law + Redfield, Eq.~\eqref{additive1}], while the violet and green dashed lines show the individual contributions from the power-law and Redfield components, respectively. \textbf{f} Muon-spin polarization curves measured at various $B_{\rm LF}$ fields as a function of the scaling variable $t/B_{\rm LF}^{0.3}$.}
\end{figure*}

Further insight into the low-$T$ magnetic ground state of YbZn$_2$GaO$_5$ is provided by muon-spin relaxation ({\textmu}SR) experiments, performed in zero magnetic field (ZF) down to 30\,mK. Implanted muons can detect internal field as low as 10\,{\textmu}T and represent highly sensitive probes of magnetic order (or the absence of it). 
Muons also give access to the dynamics 
of spin fluctuations in a frequency range complementary to that of NMR. {\textmu}SR measurements were performed on the FLAME spectrometer at the Paul Scherrer Institute on two plate-like single crystals with different dimensions: $4\times4\times0.5$\,mm$^3$ and $2\times2\times0.5$\,mm$^3$. The crystals were co-aligned in the horizontal $ab$-plane and glued on a thin copper foil using GE 7031 varnish. 

Figure~\ref{Fig1}(a) depicts a few representative muon-spin asymmetry curves measured in a broad temperature range, from 0.03\,K to 75\,K, in zero field.
Muon asymmetry spectra show a smooth, continuous decrease in polarization down to 30\,mK, with no evidence of coherent oscillations. This behavior indicates the absence of static, uniformly ordered local magnetic fields, which rules out long-range magnetic order~\cite{Amato2024}. A disordered but static spin system, such as a canonical spin glass, usually shows
a strongly damped, non-oscillatory decay, followed by a recovery to one-third of the initial asymmetry, due to the presence of frozen, randomly oriented moments. The absence of such a $1/3$-tail in our 30-mK dataset suggests the absence of 
spin freezing in this system, even at the lowest temperatures~\cite{Mendels077204, Uemura546}.

The ZF-{\textmu}SR asymmetry spectra $A_{\rm ZF}(t)$ across the entire 0.03--75\,K temperature range are consistently described by a stretched exponential function
\begin{equation}
	\label{fit_function_1}
	A_{\rm ZF/LF}(t) = A_{\rm rel} e^{-(\lambda_{\rm ZF/LF} t)^\beta} + A_{\rm b},
\end{equation}
where $\lambda$ is the temperature-dependent spin relaxation rate, $A_{\rm rel} \approx 0.17$ is the relaxing asymmetry, and $A_{\rm b} \approx 0.07$ represents the non-relaxing background contribution from muons stopping in the sample holder. Although the $A_{\rm b}$ fraction appears relatively large (i.e., well above the level typically expected on FLAME for such a large sample), it could also reflect a substantial contribution from slowly fluctuating internal fields aligned along the initial muon-spin direction, manifesting as the residual asymmetry persisting at long times. The extracted values of $\lambda$ and $\beta$ are shown in Fig.~\ref{Fig1}(b) and (c), respectively. At high temperatures ($6\,\mathrm{K} \leq T \leq 75\,\mathrm{K}$), the relaxation rate $\lambda$ is nearly temperature independent, while the stretching exponent $\beta$ remains close to unity ($0.9 \leq \beta \leq 1$), indicating near-exponential relaxation.This behavior is consistent with a thermally activated process, as the present $\mu$SR measurements probe the regime $k_{\mathrm{B}}T \leq E_A$, where the temperature dependence of $\lambda$ is weak; a clear decrease in $\lambda(T)$ appears only at higher temperatures ($T \geq 70$\,K), as established by NMR data and a previous $\mu$SR study~\cite{wu2025spin}.

In a densely distributed magnetic environment, the fluctuating local magnetic-field distribution $B_{\mathrm{loc}}$ at the muon stopping sites can be reasonably approximated by a Gaussian distribution with a width $\Delta \sim \gamma_{\mu}\sqrt{\langle (B_{\mathrm{loc}} - \langle B_{\mathrm{loc}} \rangle)^2 \rangle}$, where $\gamma_\mu = 2\pi \times 135.5~\mathrm{MHz\,T^{-1}}$ is the muon gyromagnetic ratio~\cite{Uemura3306, Li097201}. Using the reported value for the nearest-neighbor exchange coupling between the $S=1/2$ Yb$^{3+}$ magnetic moments, $J_{1}/k_{\rm B} \simeq 0.5$\,meV ($\sim 5.8$\,K)~\cite{Bag266703}, the Yb$^{3+}$ spin-fluctuation rate in the high-$T$ limit is estimated as $\nu = \sqrt{z}\,J_{1}S/\hbar \approx 9.3 \times 10^{11}~\mathrm{s^{-1}}$, where $z=6$ is the number of nearest neighbors for a triangular lattice. Within a single-time relaxation model, the longitudinal-field dependence of the relaxation rate is described by the Redfield expression~\cite{Redfield1965}:

\begin{equation}
\label{Redfield}
\lambda(B_{\mathrm{LF}}) =
\frac{2\Delta^2}{\nu}
\frac{1}{1 + \left( \frac{\omega_\mu}{\nu} \right)^2 },
\end{equation}
where $\omega_\mu = \gamma_\mu B_{\mathrm{LF}}$.

Using $\lambda \approx 0.16(1)$~$\mu\mathrm{s}^{-1}$ at $T \geq 6$~K and $B_{\mathrm{LF}}=0$, the fluctuating field distribution width is estimated to be $\Delta \sim 2.75(1)\times10^{8}~\mathrm{s^{-1}}$, validating the fast fluctuation regime $\nu \gg \Delta$.

As the temperature is lowered below $\sim 6$\,K, $\lambda$ begins to increase, indicating a gradual slowing down of magnetic fluctuations, as spin correlations develop below an energy scale comparable to the dominant exchange coupling $J_{1} \simeq 5.8$\,K. Simultaneously, the stretching exponent $\beta$ decreases, reaching a shallow minimum ($\beta \simeq 0.8$) around 2\,K, followed by a weak recovery to $\beta \simeq 0.85$ at 0.3\,K, defining a broad crossover region spanning $0.3\,\mathrm{K} \leq T \leq 6\,\mathrm{K}$. This temperature evolution of $\lambda$ and $\beta$ reflects the development of a broad distribution of relaxation rates, consistent with increasingly inhomogeneous and spatially distributed spin dynamics. Below 0.3\,K, down to the base temperature of 30\,mK, both $\lambda$ and $\beta$ saturate at approximately $\lambda \simeq 0.3$\,{\textmu}s$^{-1}$ and $\beta \simeq 0.85$. The absence of any further suppression of $\lambda$ at low temperatures argues against a fully gapped ground state and indicates a persistent, gapless spin dynamics~\cite{Mendels077204, Faak037208, Quilliam214432}.

While the overall temperature evolution of $\lambda$ is consistent with the previous $\mu$SR study~\cite{wu2025spin}, a key difference lies in the form of the relaxation. Earlier work reported a single-exponential behavior ($\beta = 1$), implying a relatively homogeneous magnetic environment, whereas our data show $\beta < 1$ at all temperatures, indicating a persistent distribution of relaxation rates. This latter distribution can be understood in terms of Zn/Ga site disorder. Disorder adjacent to the Yb--O--Yb superexchange pathways may lead to a distribution of Yb$^{3+}$ $g$-factors thus introducing exchange randomness, providing a possible microscopic origin for the distributed spin dynamics. Besides, Zn/Ga site disorder is expected to generate a continuum of locally inequivalent environments, thereby broadening the otherwise discrete set of candidate muon sites~~\cite{wu2025spin}. 
In this situation, the number of distinct relaxation channels is not uniquely defined. A stretched-exponential function therefore provides a natural phenomenological description of a broad distribution of relaxation rates. Nevertheless, we also performed a two-component analysis, presented in the Supplementary Material~\cite{SM}, as an independent check of the fitting model. This alternative parametrization reproduces the same qualitative trends and supports the same physical conclusions, indicating that our main conclusions are not tied to a specific fitting parametrization.


To further confirm the dynamic character of the ground state, we examined the longitudinal-field (LF) dependence of the muon relaxation rate at 0.1\,K. As shown in Fig.~\ref{Fig1}(d), the persistence of relaxation up to longitudinal fields of $\sim 3.5$\,T indicates that the dominant contribution arises from fluctuating electronic moments. To quantify the field dependence of the relaxation rate, we fitted the longitudinal-field asymmetry spectra $A_{\rm LF}(t)$ using the same stretched exponential form [Eq.~\eqref{fit_function_1}] as in zero field, keeping $A_0$ as a global free parameter while fixing $A_{\rm b}$ to the value determined from zero-field fit at 0.1\,K.

The LF relaxation rate $\lambda$ is directly sensitive to the low-energy spin excitation spectrum $\tilde{S}(\omega)$ and is related to the Fourier transform of the spin autocorrelation function $\langle \tilde{S}(t)\tilde{S}(0) \rangle$ via the relation:

\begin{equation}
\lambda \sim \int_0^{\infty} 
\langle \tilde{S}(t)\tilde{S}(0)\rangle
\cos(\omega_\mu t)\,\mathrm{d}t,
\end{equation}

This relationship enables us to deduce the dynamical properties of the magnetic system by examining the shape of the autocorrelation function. In systems with uniform magnetic environments, $\langle \tilde{S}(t)\tilde{S}(0) \rangle$ decays exponentially, whereas in  disordered systems one observes slower, non-exponential decays, such as a stretched exponential or a power law.

We attempted several fits of the LF dependence of the relaxation rate $\lambda(B_{\mathrm{LF}})$ using different spin autocorrelation functions [see Supplemental Material~\cite{SM} for details]. We find that a two-component additive form is the only consistent way to describe data across the entire field range: 
\begin{equation}
\label{additive1}
\lambda(B_{\mathrm{LF}}) = \lambda_1(B_{\mathrm{LF}}) + \lambda_2(B_{\mathrm{LF}}).
\end{equation}

The first component consists of a power-law form, corresponding to a spin autocorrelation function
\begin{equation}
\tilde{S}(t) \sim \left(\frac{\tau_1}{t}\right)^{x_1} e^{-\nu_1 t},
\end{equation}
capturing the slow, non-Lorentzian dynamics. Here, $\tau_1$ defines a short-time cutoff, while $\nu_1^{-1}$ sets the characteristic timescale for the long-time decay, and $x_1$ is a critical exponent characterizing the slow decay. This leads to the corresponding expression for the relaxation rate~\cite{Keren054403, Li097201}:
\begin{equation}
\lambda_1(B_{\mathrm{LF}}) =
2\Delta_1^2 \tau_1^{x_1}
\int_0^\infty \!\! t^{-x_1} e^{-\nu_1 t}
\cos(\omega_\mu t)\,\mathrm{d}t.
\label{eq:powerlaw1}
\end{equation}

The second component is modeled using the Redfield expression [Eq.~\eqref{Redfield}], corresponding to an exponential spin autocorrelation
\begin{equation}
\tilde{S}(t) \sim e^{-\nu_2 t}, 
\end{equation}
with $\Delta_2$ and $\nu_2$ representing the field distribution width and fluctuation rate, respectively. Thus, the two components describe two distinct dynamical processes: a slow, correlated channel and a faster single-timescale dynamic channel.

In Fig.~\ref{Fig1}(e), the field dependence of the relaxation rate is shown together with the corresponding fit using Eq.~\eqref{additive1}, along with the individual contributions from the two components. The power-law contribution is described by an effective prefactor $A_{\mathrm{eff}} = 2\Delta_1^2 \tau_1^{x_1}$, which sets the overall amplitude of the relaxation. The cutoff timescale is fixed to $\tau_1 = 10^{-12}$~s~\cite{Li097201}, corresponding to a microscopic picosecond timescale associated with the onset of spin correlations. Such short-time dynamics lies well outside the experimentally accessible $\mu$SR time window, and therefore $\tau_1$ cannot be independently determined from the data. As a result, $A_{\mathrm{eff}}$ should be regarded as an effective normalization parameter rather than a direct microscopic field scale. The fit yields $x_1 = 0.54(11)$ and $\nu_1 = 7.78(3.30)\times10^{6}~\mathrm{s^{-1}}$, where $\nu_1$ represents the characteristic fluctuation rate governing the long-time decay of the correlations. Both $x_1$ and $\nu_1$ primarily determine the field dependence of the power-law component. The second component yields $\Delta_2/2\pi = 0.73(11)~\mathrm{MHz}$, corresponding to an effective local field of $\sim 5.4(8)$~mT, and a fluctuation rate $\nu_2 = 4.24(0.97)\times10^{8}~\mathrm{s^{-1}}$, associated with a single-timescale, Redfield-like fluctuation channel.

The fit provides a reasonable description of the data over the full field range, with $\chi^2_{\mathrm{red}} \approx 1.2$. As evident from the individual component curves, the power-law contribution dominates in the low-field regime, while above $\sim 0.02$~T the Redfield-like component becomes progressively dominant as the power-law term is suppressed with increasing longitudinal field. The power-law form reflects slow, non-exponential dynamics. In contrast, the second component follows a conventional Redfield form and is characterized by a single fluctuation timescale $\nu_2$. The hierarchy $\nu_2 > \nu_1$ is consistent with the two-channel dynamical picture discussed above. Both fluctuation rates are significantly reduced compared to the high-temperature limit ($\sim 9.3 \times 10^{11}$~$\mathrm{s^{-1}}$), pointing to a pronounced slowing down of spin dynamics upon cooling.

Additional support for the power-law spin dynamics is provided by a time–field scaling analysis of the $A_{\mathrm{LF}}(t)$ data, which collapse onto a universal curve when plotted as a function of $t/B_{\mathrm{LF}}^{\gamma}$ with $\gamma = 0.3$ [Fig.~\ref{Fig1}~(f)]. Such scaling indicates non-exponential spin correlations arising from a broad distribution of fluctuation timescales, consistent with a distributed spin dynamics. The scaling breaks down at higher fields ($B_{\mathrm{LF}} > 0.3$~T), with clear deviations from scaling observed above $\sim 0.5$~T, reflecting the significant suppression of the power-law contribution and the increasing dominance of single-timescale (Lorentzian) dynamics, as also evident from the individual component curves shown in Fig.~\ref{Fig1}~(e), consistent with the field-induced suppression of the slower components of the distributed dynamics.

We also analyzed the field dependence of the relaxation rate using a 2D spin-diffusion model including an additional non-diffusive (0D) Redfield-like contribution, following the approach of Ref.~\cite{wu2025spin} (see Supplementary Material~\cite{SM} for details). This model provides a satisfactory description of the data, comparable to that obtained from the two-component [Eq.~\eqref{additive1}] analysis. Both approaches provide comparable descriptions of the data and are consistent with the two-channel dynamical picture discussed above. The 2D+0D model attributes the slow component to diffusive spin transport, whereas the power-law description provides a more general representation of distributed dynamics, captured by the exponent $x_{1}$. Taken together, these results indicate that the spin dynamics cannot be described by a single characteristic timescale, but instead arises from the coexistence of spatially extended, slow correlated dynamics and a faster, non-diffusive (Redfield-like) contribution.

In the present system, the observed distributed dynamics can be naturally linked to Zn/Ga site disorder, as shown by our $^{71}$Ga-NMR measurements. Consistent with DFT+$\mu$ calculations~\cite{wu2025spin}, which place the muon stopping sites near oxygen atoms bonded to Zn and Ga, the muon probes the spatial variation of the local magnetic environments. Disorder-induced variations in exchange pathways and internal magnetic fields therefore lead to a wide spread of the local environments, manifested as the power-law behavior observed in the LF-$\mu$SR data~\cite{Kermarrec100401, Keren107204, Helton147201}. This provides a natural microscopic origin for the slow, non-Lorentzian component of the relaxation, while the faster Redfield-like channel reflects more local, non-diffusive short-range fluctuations characterized by a single timescale. Together, these results highlight the inhomogeneous nature of the low-temperature spin dynamics, arising from the underlying disorder-dominated landscape.

Remarkably, while the $\mu$SR and NMR relaxation rates $\lambda$ and $1/T_1$ show similar behaviors at high temperatures, their temperature variations differ below $\sim 10$~K. The difference should primarily be attributed to strong field effects, as already pointed out in Ref.~\cite{wu2025spin} with a crossover from a decreasing to an increasing variation around 0.5~T. This suggests that the distributed and single time relaxations, dominating respectively at low and high applied fields, experience different variation with temperature. Besides, heat capacity measurements~\cite{Bag266703} also give evidence for a strong sentitivity to the applied magnetic field of the spin liquid-like ground state realized in YbZn$_2$GaO$_5$.

\section{Discussion and conclusion}
In summary, our combined NMR and $\mu$SR results support a 
spin-fluctuating ground state that persists down to at least 0.03\,K. From these results alone, it is difficult to determine the extent to which disorder influences the low-energy dynamics, as similar $\mu$SR behavior has also been observed in Yb(Mg/Zn)GaO$_4$.
They set in a context where different bulk- and local-probe measurements suggest that Yb\-Zn$_2$\-GaO$_5$ exhibits QSL behavior. Previously reported heat-capacity measurements at a zero magnetic field indicate an absence of phase transitions down to 60\,mK, as well as a quadratic temperature dependence up to 100\,mK. This suggests that the physics of the Dirac cones is being probed within this temperature range. 
The dependence of the heat capacity on the applied magnetic field further validates this interpretation, demonstrating a linear increase with the magnetic field up to 0.05\,T, thus providing an indication of the magnetic field range for which the system persists in the Dirac regime. 
Our $\mu$SR measurements likewise detect neither long-range magnetic order nor spin freezing down to 30 mK, in line with \cite{wu2025spin}. As discussed above, this temperature range lies well within the low-temperature regime relevant to our system. While recent ab initio calculations on the ideal structure predict a 120$^{\circ}$ Néel ordered ground state \cite{abinitio}, no signatures of such order are observed down to the base temperature of the $\mu$SR dilution refrigerator. Instead, a quantum Fisher information (QFI) study based on the $\mu$SR data in \cite{wu2025spin} supports a QSL phase in YbZn$_2$GaO$_5$. Indeed, QFI has recently been proposed as a sensitive entanglement-based metric to distinguish intrinsic quantum spin liquids from disorder-driven random-singlet states in triangular-lattice antiferromagnets~\cite{shimokawa2025}.  

Since our NMR results reveal a substantial site mixing between the Ga$^{3+}$ and Zn$^{2+}$ cations, clarifying the nature of the ground state becomes a central question. A similar site mixing has been reported in the sister compounds YbMgGaO$_4$ and YbZnGaO$_4$, where a spin-glass phase was identified and attributed to exchange randomness effects~\cite{Ma087201, Li107202}. In our case, however, the spin-glass scenario is ruled out by $\mu$SR (this study and \cite{wu2025spin}) as well as AC-susceptibility measurements~\cite{Bag266703}. 

Several scenarios have been discussed to understand the effect of disorder on a triangular lattice: a theoretical study on an anisotropic $J_1$-$J_2$ $XXZ$ model, with random pseudo-dipolar terms, shows that disorder can induce a phase with no long-range order, which exhibits a broad INS continuum reminiscent of a QSL~\cite{Zhu157201}. A more recent study, using quantum Fisher information on a disordered $J_1$-$J_2$ $XXZ$ Heisenberg model, points in a similar direction \cite{shimokawa2025}. 
Nevertheless, the nature of the induced phase differs fundamentally from that of a quantum spin liquid. In the former case, it is a superposition of different stripe phases; in the latter, it is a random-singlet phase. 
On the other hand, another theoretical study on a $J_1$-$J_2$ $XXZ$ Heisenberg model revealed  that bond randomness can enlarge the stability region of an intrinsic QSL phase and induce a QSL phase in systems that would otherwise be magnetically ordered~\cite{wu2019randomness}.
In the YbZn$_2$GaO$_5$ case, no comprehensive combined theoretical and experimental effort has yet determined the appropriate Hamiltonian, the relevant exchange couplings, or the quantitative role of disorder on the Yb planes.

As a result, positioning this material within the phase diagrams proposed by the aforementioned theoretical studies remains an interesting issue. Experimentally, site mixing that is now clearly established and the spin liquid fingerprints from heat capacity and $\mu$SR are prominent constraints for further models. We note that a similar heat-capacity signature has been reported in the kagomé system YCu$_3$(OH)$_6$Br$_2$[Br$_x$(OH)$_{1-x}$] where chemical disorder suppresses the magnetic order present in the ordered Y$_3$Cu$_9$(OH)$_{19}$Cl$_8$ parent compound~\cite{Chatterjee2023}, and stabilizes a possible Dirac QSL phase \cite{PhysRevB.105.L121109}. Whether the behavior observed in YbZn$_2$GaO$_5$ is likewise disorder-induced remains an open question requiring further investigation.

\subsection{Acknowledgement}

We acknowledge the support of the French Agence Nationale de la Recherche, under Grant No. ANR-25-CE30-2010-01 “ULTIMAT”, the Fondation Charles Defforey-Institut de France and the program France 2030 (QuanTEdu-France, ANR-22-CMAS-0001). Work performed at Duke University is supported by the U.S. Department of Energy, Office of Science, Office of Basic Energy Sciences, under Award No. DE-SC0023405. 

\subsection{Data availability}

The data that support the findings of this article are not publicly available upon publication because it is not technically feasible and/or the cost of preparing, depositing,
and hosting the data would be prohibitive within the terms of this research project. The data are available from the authors upon reasonable request.


%

\clearpage
\onecolumngrid

\setcounter{section}{0}
\setcounter{subsection}{0}
\setcounter{figure}{0}
\setcounter{table}{0}
\setcounter{equation}{0}
\renewcommand{\thetable}{S\arabic{table}}
\renewcommand{\thefigure}{S\arabic{figure}}
\renewcommand{\theHsection}{supp.\arabic{section}}
\renewcommand{\theHsubsection}{supp.\arabic{section}.\arabic{subsection}}
\renewcommand{\theHfigure}{supp.\arabic{figure}}
\renewcommand{\theHtable}{supp.\arabic{table}}
\renewcommand{\theHequation}{supp.\arabic{equation}}

\begin{center}
{\Large\bfseries Supplemental Material for ``Interplay of disorder and spin liquid physics in the triangular-lattice antiferromagnet YbZn$_2$GaO$_5$''}\par
\vspace{1em}
{\normalsize
Hafsa Zeroual,$^{1,*}$ Shams Sohel Islam,$^{2,*}$ Toni Shiroka,$^{2}$ Thomas James Hicken,$^{2}$
Rabindranath Bag,$^{3}$ Sara Haravifard,$^{3}$ Hubertus Luetkens,$^{2}$ Philippe Mendels,$^{1}$
Edwin Kermarrec,$^{1}$ and Fabrice Bert$^{1, \dagger}$}\par
\vspace{0.7em}
{\small
$^{1}$Universit\'e Paris-Saclay, CNRS, Laboratoire de Physique des Solides, 91405, Orsay, France\\
$^{2}$PSI Center for Neutron and Muon Sciences CNM, 5232 Villigen PSI, Switzerland\\
$^{3}$Department of Physics, Duke University, Durham, NC, USA\\
$^{*}$These authors contributed equally to this work.\\
$^{\dagger}$fabrice.bert@universite-paris-saclay.fr}
\end{center}

\vspace{1em}
These Supplementary Materials provide additional experimental details, analysis procedures, and supporting data for the main manuscript. They include additional figures, and further discussion of data analysis.

\section{NMR}
	\subsection{Knight shift analysis}
    	\begin{figure}[hbtp]
		\centering
		\includegraphics[width=\linewidth]{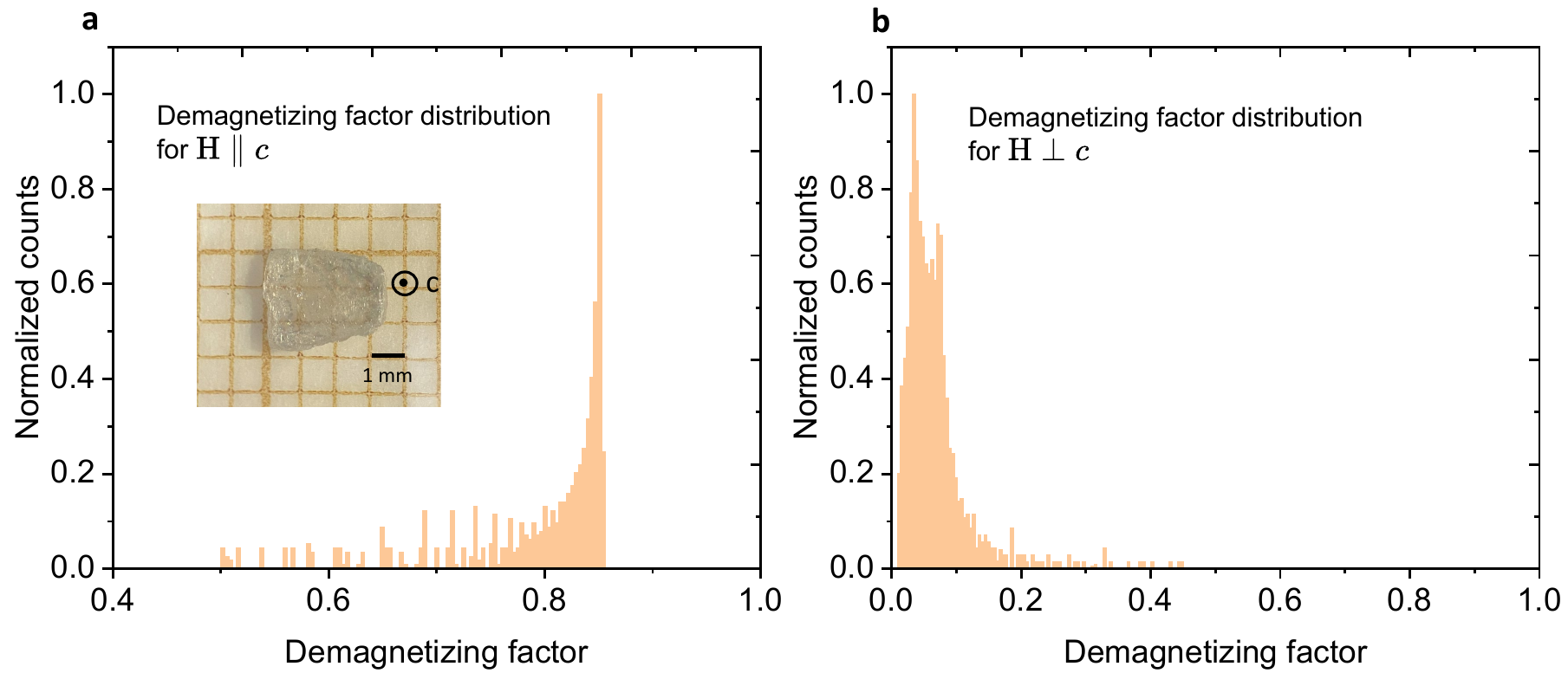}
		\caption{Distribution of the demagnetization factor for our NMR sample for the two field orientations $\mathbf{H} \parallel c$ and $\mathbf{H} \perp  c$.}
		\label{fig:demag}
	    \end{figure}
    Our NMR measurements were performed on an almost rectangular single crystal, with the largest facet ($\sim 3 \times 3$~mm$^2$) corresponding to the $ab$ plane and the shortest dimension ($\sim 0.5$~mm) along the $c$ axis, as shown in Fig.~~\ref{fig:demag}. As this sample geometry significantly deviates from a spherical shape, corrections due to the demagnetizing field contribution to the NMR shift are required for both field orientations, $\mathbf{H} \parallel c$ and $\mathbf{H} \perp c$.
	
	In a magnetic material, the demagnetizing field can be expressed as
	\begin{equation}
		H_i^{\mathrm{demag}} = - \sum_j N_{ij} M_j,
	\end{equation}
	where $H_i^{\mathrm{demag}}$ is the $i$-th component of the demagnetizing field, $M_j$ is the $j$-th component of the magnetization, and $N_{ij}$ is the demagnetization tensor.

	We numerically evaluated the distribution of demagnetization factors for our sample geometry following Ref.~\cite{wysin2012demag}. The magnetization is assumed to be uniform and collinear with the applied magnetic field. Figure.~\ref{fig:demag} shows the resulting distribution of the relevant demagnetization factor for both field orientations.

	We focus on the diagonal component $N_{uu}$, where $u$ denotes the direction of the applied magnetic field, as the off-diagonal components average to zero for a uniform magnetization. The mean values are found to be $N_{uu} = -0.069$ for $\mathbf{H} \perp c$ and $N_{uu} = -0.789$ for $\mathbf{H} \parallel c$.
	
	Using these values, the corresponding demagnetizing fields for a magnetic moment of $1~\mu_B$ per $\mathrm Yb^{3+}$ are estimated as
	\begin{equation}
		B_{\mathrm{demag}} = \mu_0 H^{\mathrm{demag}} = -\mu_0 N_{uu} M,
	\end{equation}
	yielding $B_{\mathrm{demag}}^{H \parallel ab} \approx  -7~\mathrm{mT}$ and $B_{\mathrm{demag}}^{H \parallel c} \approx -85~\mathrm{mT}$.
	
\begin{figure}[hbtp]
		\centering
		\includegraphics[width=\linewidth]{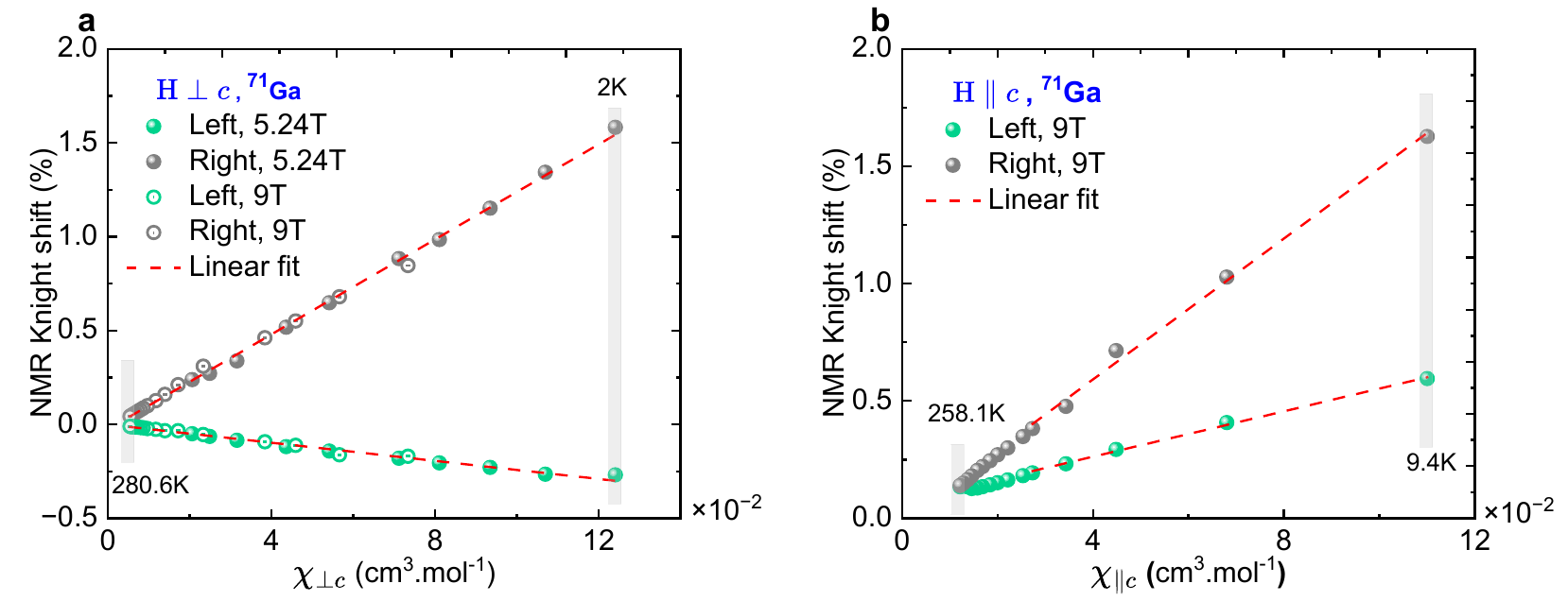}
		\caption{$K$--$\chi$ for both field directions and sites.}
		\label{fig: shift}
		\end{figure}

	The hyperfine coupling constants extracted from the $K$ vs $\chi$ analysis are corrected for this contribution according to
	\begin{equation}
		A^{\mathrm{measured}} = A^{\mathrm{corrected}} + B_{\mathrm{demag}},
	\end{equation}
	where $A^{\mathrm{measured}}$ is the slope obtained from the $K$--$\chi$ plot and $A^{\mathrm{corrected}}$ is the corrected hyperfine coupling constant, both expressed in units of $\mathrm{mT}/\mu_B$. Fig.~\ref{fig: shift} shows the K--$\chi$ curves for both sites and both field orientations.  Table.~\ref{tab:hyperfine} summarizes the hyperfine coupling constants extracted from the linear fits together with their corrected values accounting for demagnetizing field contributions.\\

    \begin{table}[htbp]
	\centering
	
	\begin{tabular}{c c c c c}
		\hline\hline
		Site & $A^{\mathrm{measured}}_{\perp c}$ & $A^{\mathrm{corrected}}_{\perp c}$ &
		$A^{\mathrm{measured}}_{\parallel c}$ & $A^{\mathrm{corrected}}_{\parallel c}$ \\
		\hline
		Left peak  & -14 & -6 & 27 & 112 \\
		Right peak & 71 & 78 & 84 & 169\\
		\hline\hline
	\end{tabular}
	\caption{Hyperfine coupling constants $A$ (in $\mathrm{mT}/\mu_B$) before and after demagnetization correction.}
	\label{tab:hyperfine}
	\end{table}

\newpage
    
    \subsection{$\mathrm T_1$ Measurements}
	\begin{figure}[hbtp]
	\centering
	\includegraphics[width=\linewidth]{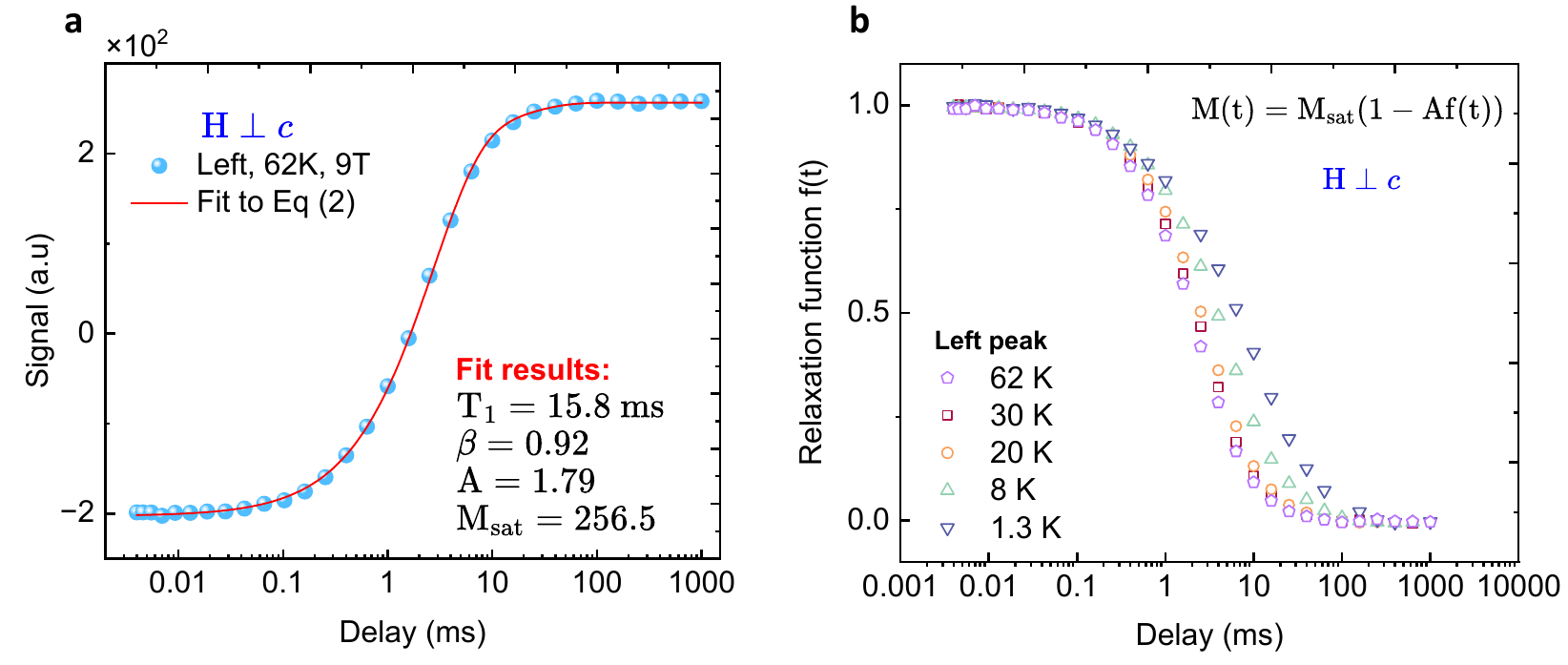}
	\caption{\textbf{a} $T_1$ recovery curve (raw data) measured using a saturation–recovery sequence. The red line shows the fit to Eq.~(2).
    \textbf{b} Superposition of the experimental relaxation functions f(t)  at different temperatures and fields ($9$ T for the curve at $62$ K and $5.24$ T for the curves $1.3$ K-- $30$ K).}
	\label{fig:recovery}
	\end{figure}

Figure~\ref{fig:recovery}{\bf .a} shows a representative example of a recovery curve measured at 9~T and 62~K together with the fit to Eq.~(2) providing the corresponding $T_1$ value. As shown on panel {\bf b}, the relaxation shape evolves at lower temperatures signaling both the increase of the average $T_1$ and of its distribution width (reduction of $\beta$). To facilitate the comparison, panel {\bf b} shows the relaxation function $f(t)=(1-M(t)/M_{sat})/A$, which is independent of the experimental conditions.

    \begin{figure}[hbtp]
	\centering
	\includegraphics[width=0.5\linewidth]{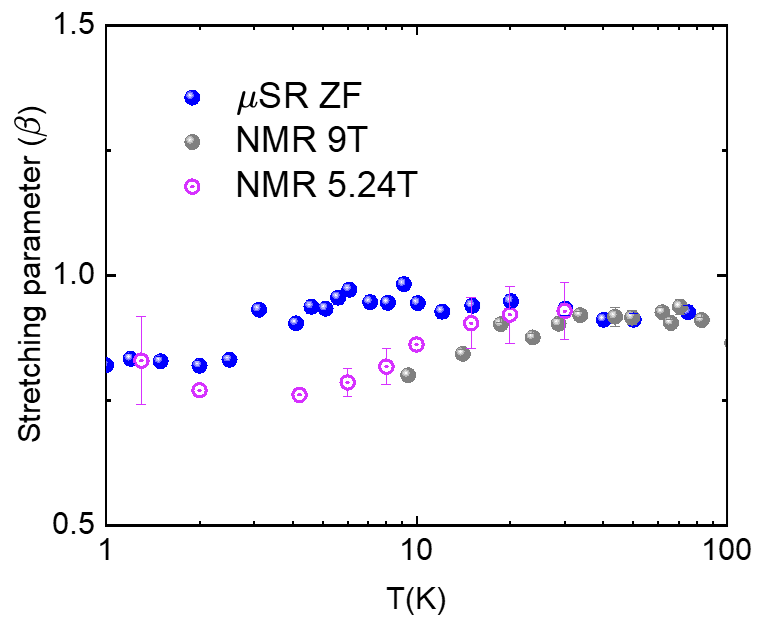}
	\caption{Comparison to the temperature evolution of the stretching exponents $\beta_{\mathrm{NMR}}$ and $\beta_{\mu \mathrm{SR}}$.} 
	\label{fig:beta}
	\end{figure}

Figure~\ref{fig:beta} shows the evolution of the stretching parameters $\beta_{\mathrm{NMR}}$ and $\beta_{\mu \mathrm{SR}}$ characterizing the widths of the relaxation rate distributions measured respectively with NMR [Eq.~(2)] and $\mu$SR [Eq.~(4)].

In the temperature range $15$--$75$~K, both stretching exponents take comparable values $\beta \simeq 0.93$ before decreasing to $\simeq 0.8$ at lower temperatures. However, $\beta_{\mathrm{NMR}}$ begins to decrease below $\sim 15$~K, while $\beta_{\mu\mathrm{SR}}$ remains nearly constant down to $\sim 3$~K before decreasing. The earlier reduction of $\beta_{\mathrm{NMR}}$ upon cooling may be attributed to an imperfect treatment of the increasing contribution of the quadrupolar signal from the secondary site and/or the overlap of the central lines, which lead to an apparent broadening of the relaxation-time distribution. In this respect, the $T$ evolution of $\beta_{\mu\mathrm{SR}}$ is likely more reliable.

\FloatBarrier

\section{$\mu$SR}

\subsection{ZF-$\mu$SR data analysis using a double exponential model}

\begin{figure*}
\centering
	\includegraphics[width = 0.6\linewidth]{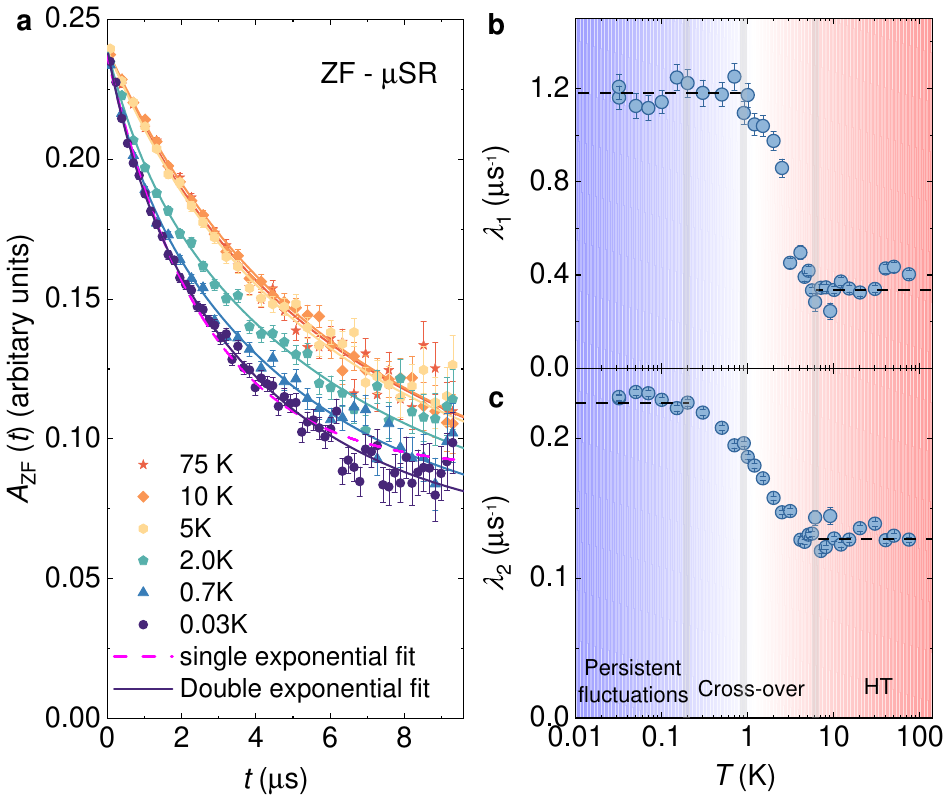}
	\caption{\label{FigA2} 
	\textbf{a} Temperature evolution of the zero-field (ZF) muon spin asymmetry, $A_{\rm ZF}(t)$, for YbZn$_2$GaO$_5$. The solid lines represent fits to the data using a double-exponential relaxation function [Eq.~\eqref{doubleExpo}]. The temperature dependence of the corresponding relaxation rates is shown in \textbf{b} $\lambda_{1}$ and \textbf{c} $\lambda_{2}$, representing the early and late-time relaxing components, respectively.}
\end{figure*}

Figure~\ref{FigA2}\,\textbf{a} shows representative zero-field (ZF) $\mu$SR asymmetry spectra collected over the temperature range $0.03$--$75$~K. Throughout this range, the asymmetry exhibits a smooth, monotonic relaxation without any anomalous features. The data at all temperatures can be consistently described using a two-component exponential form,
\begin{equation}
\label{doubleExpo}
A_{\mathrm{ZF}}(t) = A_{\mathrm{rel}} \left[ f\, e^{-\lambda_1 t} + (1 - f)\, e^{-\lambda_2 t} \right] + A_b,
\end{equation}
where $\lambda_1$ and $\lambda_2$ primarily govern the short- and long-time evolution of $A_{\mathrm{ZF}}(t)$, respectively, and $f$ represents the relative weight of the faster relaxing component. A satisfactory description is obtained by fixing $f \approx 0.19$ across the full measured temperature range. The relaxing asymmetry is $A_{\mathrm{rel}} \approx 0.17$, while a small, temperature-independent background $A_{\rm b} \approx 0.07$ arises from muons stopping outside the sample (e.g., cryostat walls or sample holder).\\
As shown in Fig.~\ref{FigA2}\,\textbf{b-c}, at high temperatures ($6\,\mathrm{K} \leq T \leq 75\,\mathrm{K}$), both $\lambda_1$ and $\lambda_2$ remain nearly temperature independent. Upon cooling below $\sim 6$~K, both relaxation rates increase, reflecting a progressive slowing down of the spin dynamics. While the overall temperature dependence is qualitatively similar, $\lambda_1$ tends to saturate below $\sim 1$~K, whereas $\lambda_2$ continues to evolve down to lower temperatures and saturates only below $\sim 0.2$~K.\\
We emphasize that the use of two exponential components should be regarded as a phenomenological representation of the relaxation rather than direct evidence for two well-defined muon stopping sites. In the previous study, DFT+$\mu$ calculations identified multiple candidate muon sites, but the ZF-$\mu$SR response was argued to be dominated by one lowest-energy site, resulting in an effectively single-component relaxation~\cite{wu2025spin}. In the presence of Zn/Ga site disorder, however, the local magnetic environments experienced by the muon are expected to broaden into a distribution of inequivalent configurations. In such a scenario, the separation into two components can be viewed as an effective approximation to a continuum of relaxation channels, whereas a stretched-exponential form provides a natural description of the underlying distributed dynamics. 


\subsection{LF-$\mu$SR data analysis}

\begin{figure*}
	\includegraphics[width = \linewidth]{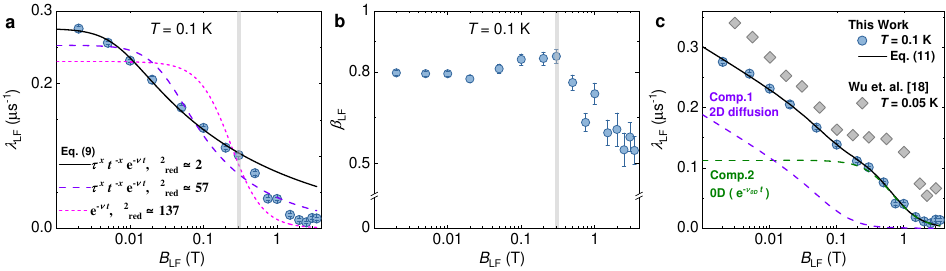}
	\caption{\label{FigA1} \textbf{a} Longitudinal-field (LF) dependence of the muon-spin relaxation rate $\lambda_{\rm LF}$. The solid black line represents a single-component fit in the restricted range ($0 \leq B_{\rm LF} \leq 0.3$~T) using the power-law form [Eq.~\eqref{supp:eq:powerlaw1}], while the long-dashed violet and short-dashed magenta lines show the fits using the power-law and conventional Redfield models, respectively. \textbf{b} Longitudinal-field dependence of the stretched exponent $\beta_{\rm LF}$, exhibiting a clear change in slope near $B_{\rm LF} \simeq 0.3$~T. \textbf{c} Longitudinal-field dependence of $\lambda_{\rm LF}$ modeled using the two-component 2D+0D framework [Eq.~\eqref{2D+0D}], together with previously reported data for $\lambda_{\rm LF}$ from Ref.~\cite{wu2025spin} for comparison.}
\end{figure*}

We first analyze the longitudinal-field (LF) dependence of the relaxation rate $\lambda(B_{\mathrm{LF}})$ using the conventional Redfield form,
\begin{equation}
\label{supp:Redfield}
\lambda(B_{\mathrm{LF}}) =
\frac{2\Delta^2}{\nu}
\frac{1}{1 + \left( \frac{\omega_\mu}{\nu} \right)^2 },
\end{equation}
where $\omega_\mu = \gamma_\mu B_{\mathrm{LF}}$, $\gamma_\mu = 2\pi \times 135.5~\mathrm{MHz\,T^{-1}}$, 
which assumes an exponentially decaying spin autocorrelation function,
\begin{equation}
\tilde{S}(t) \sim e^{-\nu t}, 
\end{equation}
As shown by a short-dashed magenta line in Fig.~\ref{FigA1}\,\textbf{a}, this model fails to reproduce the experimental data over the full field range, indicating that a single-timescale description is insufficient.

We next model $\lambda(B_{\mathrm{LF}})$ using a generalized power-law form~\cite{Keren054403, Li097201},
\begin{equation}
\lambda_1(B_{\mathrm{LF}}) =
2\Delta^2 \tau^{x}
\int_0^\infty \!\! t^{-x} e^{-\nu t}
\cos(\omega_\mu t)\,\mathrm{d}t.
\label{supp:eq:powerlaw1}
\end{equation}
corresponding to a non-exponential spin autocorrelation function of the form, 
\begin{equation}
\tilde{S}(t)\sim t^{-x}e^{-\nu t}. 
\end{equation}
In this formulation, $\Delta$ and $\tau$ enter only through the effective prefactor $A_{\mathrm{eff}}=2\Delta^2\tau^{x}$, which primarily sets the overall magnitude of the relaxation rate, whereas the field dependence is governed by $x$ and $\nu$. We therefore treat $A_{\mathrm{eff}}$ as an effective normalization factor and fix $\tau=10^{-12}$~s, corresponding to a microscopic picosecond cutoff timescale~\cite{Li097201}. We have verified that varying $\tau$ within a reasonable microscopic range does not affect the extracted values of $x$ and $\nu$, but only rescales the prefactor $A_{\mathrm{eff}}$, indicating that the physical conclusions are insensitive to the precise choice of $\tau$. As shown in Fig.~\ref{FigA1}\,\textbf{a}, this single power-law form provides a good description in the low-field regime ($B_{\mathrm{LF}}\leq0.30$~T), yielding $x=0.79(3)$ and $\nu=5.4(2.0)\times10^{6}~\mathrm{s^{-1}}$, with $\chi^2_{\mathrm{red}}\approx2.09$. However, when extended over the full field range, the same model fails to reproduce the high-field behavior, yielding $x=0.58(12)$ and $\nu=2.5(1.9)\times10^{7}~\mathrm{s^{-1}}$, with a significantly larger $\chi^2_{\mathrm{red}}\approx57.1$.

These results demonstrate that a single dynamical channel, even with a non-exponential power-law autocorrelation function, is insufficient to describe the full field dependence of $\lambda(B_{\mathrm{LF}})$. This conclusion is further supported by the field dependence of the stretched exponent $\beta_{\rm LF}$ [Fig.~\ref{FigA1}\,\textbf{b}], which exhibits a clear change in slope around $0.3$~T, indicating a crossover in the underlying relaxation mechanism.

We therefore consider a two-component description combining a power-law and a conventional Redfield contribution, as introduced in the main text. In this framework, the power-law term captures slow, distributed dynamics characterized by a broad distribution of fluctuation timescales, while the Redfield term describes a faster single-timescale relaxation channel. Here, the two contributions are treated additively to model the full field dependence of $\lambda(B_{\mathrm{LF}})$, allowing a comparison with the alternative 2D+0D description discussed below.

An alternative description is provided by the 2D+0D model~\cite{wu2025spin}, in which the LF relaxation is decomposed into a two-dimensional (2D) diffusive contribution and a non-diffusive zero-dimensional (0D) Redfield-like term,
\begin{equation}
\label{2D+0D}
\lambda(B_{\mathrm{LF}})=
\lambda_{\mathrm{2D}}(B_{\mathrm{LF}})
+\lambda_{\mathrm{0D}}(B_{\mathrm{LF}}).
\end{equation}

The 2D contribution is expressed as
\begin{equation}
\lambda_{\mathrm{2D}}(B_{\mathrm{LF}})
=
\frac{A_{\mathrm{2D}}^{2}}{4}
J_{\mathrm{2D}}(D_{\mathrm{2D}},\omega_e),
\end{equation}
where $\omega_e=\gamma_e B_{\mathrm{LF}}$, with $\gamma_e = 2\pi \times 28.024~\mathrm{GHz\,T^{-1}}$ the electron gyromagnetic ratio. The spectral density $J_{\mathrm{2D}}$ is given by the Fourier transform of the spin autocorrelation function,
\begin{equation}
J_{\mathrm{2D}}(D_{\mathrm{2D}},\omega_e)
=
2 \int_{0}^{\infty}
\cos(\omega_e t)\, P_{2D}(t)\, dt,
\end{equation}
where, for two-dimensional diffusive dynamics on a lattice,
\begin{equation}
P_{2D}(t)
=
\left[e^{-2 D_{\mathrm{2D}} t}\, I_0\!\left(2 D_{\mathrm{2D}} t\right)\right]^2,
\end{equation}
with $I_0$ the zeroth-order modified Bessel function of the first kind and $D_{\mathrm{2D}}$ the diffusion rate.
Here, $A_{\mathrm{2D}}$ denotes the effective hyperfine coupling between the muon and the electronic spins participating in the diffusive dynamics, typically mediated via a bridging oxygen. Based on DFT+$\mu$ calculations in Ref.~\cite{wu2025spin}, the muon stopping sites are located near oxygen atoms, such that the muon couples to several nearby Yb ions at different distances and through distinct bonding pathways. The 2D contribution is attributed to interaction with the Yb ion exhibiting the strongest contact hyperfine coupling along such a pathway.

The second contribution is described by a 0D Redfield-like term,
\begin{equation}
\lambda_{\mathrm{0D}}(B_{\mathrm{LF}})
=
\frac{\Delta_{\mathrm{0D}}^{2}}{4}
\frac{2/\nu_{\mathrm{0D}}}
{1+\left(\gamma_{\mu}B_{\mathrm{LF}}/\nu_{\mathrm{0D}}\right)^m},
\end{equation}
where $\Delta_{\mathrm{0D}}$ is the local field scale at the muon site and $\nu_{\mathrm{0D}}$ is the corresponding fluctuation rate. Within the same framework, this term is associated with interaction with a different neighboring Yb ion for which the coupling to the muon is predominantly dipolar.

Applying this model to the present data yields $A_{\mathrm{2D}} = 52(14)$~MHz and $D_{\mathrm{2D}} = 3.6(1.9)$~ns$^{-1}$, together with $\Delta_{\mathrm{0D}}/2\pi = 1.89(12)$~MHz and $\nu_{\mathrm{0D}} = 0.65(14)$~ns$^{-1}$, with $m=2$ fixed. This provides a good description of the data over the full field range, with $\chi^2_{\mathrm{red}} \approx 0.92$. Physically, the 2D contribution reflects spatially extended spin transport within the magnetic layers, whose weak field dependence arises from a broad distribution of correlation times characteristic of diffusive dynamics, whereas the 0D term corresponds to a faster, non-diffusive Redfield-like fluctuation channel. Overall, this analysis supports the coexistence of slow, spatially extended dynamics and a faster fluctuation component, consistent with the conclusions drawn from the two-component power-law + Redfield analysis presented in the main text.

Previously, the analysis of the reported data in Ref.~\cite{wu2025spin}, using a similar 2D+0D model (including additional background and level-crossing resonance (LCR) terms), yielded the following parameters: $A_{\mathrm{2D}} = 63$~MHz and $D_{\mathrm{2D}} = 2$~ns$^{-1}$ for the diffusive component, together with $\Delta_{\mathrm{0D}} = 18.4$~MHz and $\nu_{\mathrm{0D}} = 1.1$~ns$^{-1}$ for the 0D contribution, with $m = 7$.

The 2D channel parameters are broadly comparable between the two analyses [present work: $A_{\mathrm{2D}} = 52(14)$~MHz, $D_{\mathrm{2D}} = 3.6(1.9)$~ns$^{-1}$; Ref.~\cite{wu2025spin}: $A_{\mathrm{2D}} = 63$~MHz, $D_{\mathrm{2D}} = 2$~ns$^{-1}$], while the main differences arise in the 0D contribution [present work: $\Delta_{\mathrm{0D}} = 1.89(12)$~MHz, $\nu_{\mathrm{0D}} = 0.65(14)$~ns$^{-1}$; Ref.~\cite{wu2025spin}: $\Delta_{\mathrm{0D}} = 18.4$~MHz, $\nu_{\mathrm{0D}} = 1.1$~ns$^{-1}$]. In particular, Ref.~\cite{wu2025spin} reports a significantly larger 0D field-distribution width and a faster $\nu_{\mathrm{0D}}$, indicating a stronger and more rapidly fluctuating dipolar-coupled contribution. However, the larger 0D field-distribution width does not necessarily imply stronger disorder; it may instead reflect a more strongly coupled and well-defined dipolar channel associated with a specific muon stopping environment. This interpretation is consistent with the more pronounced high-field shoulder observed in their LF data [see Fig.~\ref{FigA1}~\textbf{c}], whereas the smoother field dependence in the present data suggests a weaker or more broadly distributed contribution. Although disorder can broaden the distribution of local fields by introducing a range of local environments and coupling strengths, the present comparison alone does not allow a definitive assessment of relative disorder levels, particularly given differences in the fitting models (e.g., inclusion of background and LCR terms), which can redistribute spectral weight among the fitted parameters. Thus, the observed differences are more likely to reflect variations in the muon stopping environment and coupling geometry, as well as differences in the relative weight of competing relaxation channels, rather than providing direct evidence for differing disorder strengths.

The exponent $m$ associated with the 0D contribution further highlights the difference between the two datasets. The value $m=7$ reported in Ref.~\cite{wu2025spin} indicates a more complex, non-exponential form of dynamics, whereas the present data are well described by $m=2$, corresponding to the conventional Redfield limit associated with an exponential field autocorrelation.

Both the power-law+Redfield and 2D+0D models yield comparable fit quality over the full field range, with $\chi^2_{\mathrm{red}} \simeq 0.92$--$1.2$. 
Slight deviations of the last two high-field data points from fit 
may reflect that an additional relaxation channel is relevant at higher longitudinal fields. One possible origin is a level-crossing resonance (LCR), as reported in YbZn$_2$GaO$_5$~\cite{wu2025spin}, where an LCR contribution centered near $2.7$~T was identified from a global multi-temperature analysis. In the present case, however, the limited number of high-field data points and the single-temperature dataset prevent a reliable inclusion of such a term, and it is therefore not pursued.

Despite differences in microscopic interpretation, both approaches provide comparable descriptions of the data and are consistent with the two-channel dynamical picture discussed in the main text. The 2D+0D model attributes the slow component to diffusive spin transport, whereas the power-law description captures distributed dynamics in a more general manner. These results indicate that the spin dynamics involve coexisting slow, spatially extended correlations and a faster single-timescale (Redfield-like) contribution.


\begin{thebibliography}{41}%
\makeatletter
\providecommand \@ifxundefined [1]{%
 \@ifx{#1\undefined}
}%
\providecommand \@ifnum [1]{%
 \ifnum #1\expandafter \@firstoftwo
 \else \expandafter \@secondoftwo
 \fi
}%
\providecommand \@ifx [1]{%
 \ifx #1\expandafter \@firstoftwo
 \else \expandafter \@secondoftwo
 \fi
}%
\providecommand \natexlab [1]{#1}%
\providecommand \enquote  [1]{``#1''}%
\providecommand \bibnamefont  [1]{#1}%
\providecommand \bibfnamefont [1]{#1}%
\providecommand \citenamefont [1]{#1}%
\providecommand \href@noop [0]{\@secondoftwo}%
\providecommand \href [0]{\begingroup \@sanitize@url \@href}%
\providecommand \@href[1]{\@@startlink{#1}\@@href}%
\providecommand \@@href[1]{\endgroup#1\@@endlink}%
\providecommand \@sanitize@url [0]{\catcode `\\12\catcode `\$12\catcode
  `\&12\catcode `\#12\catcode `\^12\catcode `\_12\catcode `\%12\relax}%
\providecommand \@@startlink[1]{}%
\providecommand \@@endlink[0]{}%
\providecommand \url  [0]{\begingroup\@sanitize@url \@url }%
\providecommand \@url [1]{\endgroup\@href {#1}{\urlprefix }}%
\providecommand \urlprefix  [0]{URL }%
\providecommand \Eprint [0]{\href }%
\providecommand \doibase [0]{https://doi.org/}%
\providecommand \selectlanguage [0]{\@gobble}%
\providecommand \bibinfo  [0]{\@secondoftwo}%
\providecommand \bibfield  [0]{\@secondoftwo}%
\providecommand \translation [1]{[#1]}%
\providecommand \BibitemOpen [0]{}%
\providecommand \bibitemStop [0]{}%
\providecommand \bibitemNoStop [0]{.\EOS\space}%
\providecommand \EOS [0]{\spacefactor3000\relax}%
\providecommand \BibitemShut  [1]{\csname bibitem#1\endcsname}%
\let\auto@bib@innerbib\@empty
\bibitem [{\citenamefont {Broholm}\ \emph {et~al.}(2020)\citenamefont
  {Broholm}, \citenamefont {Cava}, \citenamefont {Kivelson}, \citenamefont
  {Nocera}, \citenamefont {Norman},\ and\ \citenamefont
  {Senthil}}]{Broholm2020}%
  \BibitemOpen
  \bibfield  {author} {\bibinfo {author} {\bibfnamefont {C.}~\bibnamefont
  {Broholm}}, \bibinfo {author} {\bibfnamefont {R.~J.}\ \bibnamefont {Cava}},
  \bibinfo {author} {\bibfnamefont {S.~A.}\ \bibnamefont {Kivelson}}, \bibinfo
  {author} {\bibfnamefont {D.~G.}\ \bibnamefont {Nocera}}, \bibinfo {author}
  {\bibfnamefont {M.~R.}\ \bibnamefont {Norman}},\ and\ \bibinfo {author}
  {\bibfnamefont {T.}~\bibnamefont {Senthil}},\ }\bibfield  {title} {\bibinfo
  {title} {Quantum spin liquids},\ }\href
  {https://doi.org/10.1126/science.aay0668} {\bibfield  {journal} {\bibinfo
  {journal} {Science}\ }\textbf {\bibinfo {volume} {367}},\ \bibinfo {pages}
  {eaay0668} (\bibinfo {year} {2020})}\BibitemShut {NoStop}%
\bibitem [{\citenamefont {Anderson}(1973)}]{Anderson1973}%
  \BibitemOpen
  \bibfield  {author} {\bibinfo {author} {\bibfnamefont {P.~W.}\ \bibnamefont
  {Anderson}},\ }\bibfield  {title} {\bibinfo {title} {Resonating valence
  bonds: A new kind of insulator?},\ }\href
  {https://www.sciencedirect.com/science/article/pii/0025540873901670}
  {\bibfield  {journal} {\bibinfo  {journal} {Mater. Res. Bull.}\ }\textbf
  {\bibinfo {volume} {8}},\ \bibinfo {pages} {153} (\bibinfo {year}
  {1973})}\BibitemShut {NoStop}%
\bibitem [{\citenamefont {Singh}\ and\ \citenamefont {Huse}(1992)}]{Singh1992}%
  \BibitemOpen
  \bibfield  {author} {\bibinfo {author} {\bibfnamefont {R.~R.~P.}\
  \bibnamefont {Singh}}\ and\ \bibinfo {author} {\bibfnamefont {D.~A.}\
  \bibnamefont {Huse}},\ }\bibfield  {title} {\bibinfo {title}
  {Three-sublattice order in triangular- and kagom\'e-lattice spin-half
  antiferromagnets},\ }\href {https://doi.org/10.1103/PhysRevLett.68.1766}
  {\bibfield  {journal} {\bibinfo  {journal} {Phys. Rev. Lett.}\ }\textbf
  {\bibinfo {volume} {68}},\ \bibinfo {pages} {1766} (\bibinfo {year}
  {1992})}\BibitemShut {NoStop}%
\bibitem [{\citenamefont {Capriotti}\ \emph {et~al.}(1999)\citenamefont
  {Capriotti}, \citenamefont {Trumper},\ and\ \citenamefont
  {Sorella}}]{Capriotti1999}%
  \BibitemOpen
  \bibfield  {author} {\bibinfo {author} {\bibfnamefont {L.}~\bibnamefont
  {Capriotti}}, \bibinfo {author} {\bibfnamefont {A.~E.}\ \bibnamefont
  {Trumper}},\ and\ \bibinfo {author} {\bibfnamefont {S.}~\bibnamefont
  {Sorella}},\ }\bibfield  {title} {\bibinfo {title} {Long-range {N}\'eel order
  in the triangular {H}eisenberg model},\ }\href
  {https://doi.org/10.1103/PhysRevLett.82.3899} {\bibfield  {journal} {\bibinfo
   {journal} {Phys. Rev. Lett.}\ }\textbf {\bibinfo {volume} {82}},\ \bibinfo
  {pages} {3899} (\bibinfo {year} {1999})}\BibitemShut {NoStop}%
\bibitem [{\citenamefont {Zhu}\ \emph {et~al.}(2018)\citenamefont {Zhu},
  \citenamefont {Maksimov}, \citenamefont {White},\ and\ \citenamefont
  {Chernyshev}}]{Zhu2018}%
  \BibitemOpen
  \bibfield  {author} {\bibinfo {author} {\bibfnamefont {Z.}~\bibnamefont
  {Zhu}}, \bibinfo {author} {\bibfnamefont {P.~A.}\ \bibnamefont {Maksimov}},
  \bibinfo {author} {\bibfnamefont {S.~R.}\ \bibnamefont {White}},\ and\
  \bibinfo {author} {\bibfnamefont {A.~L.}\ \bibnamefont {Chernyshev}},\
  }\bibfield  {title} {\bibinfo {title} {Topography of spin liquids on a
  triangular lattice},\ }\href {https://doi.org/10.1103/PhysRevLett.120.207203}
  {\bibfield  {journal} {\bibinfo  {journal} {Phys. Rev. Lett.}\ }\textbf
  {\bibinfo {volume} {120}},\ \bibinfo {pages} {207203} (\bibinfo {year}
  {2018})}\BibitemShut {NoStop}%
\bibitem [{\citenamefont {Maksimov}\ \emph {et~al.}(2019)\citenamefont
  {Maksimov}, \citenamefont {Zhu}, \citenamefont {White},\ and\ \citenamefont
  {Chernyshev}}]{Maksimov2019}%
  \BibitemOpen
  \bibfield  {author} {\bibinfo {author} {\bibfnamefont {P.~A.}\ \bibnamefont
  {Maksimov}}, \bibinfo {author} {\bibfnamefont {Z.}~\bibnamefont {Zhu}},
  \bibinfo {author} {\bibfnamefont {S.~R.}\ \bibnamefont {White}},\ and\
  \bibinfo {author} {\bibfnamefont {A.~L.}\ \bibnamefont {Chernyshev}},\
  }\bibfield  {title} {\bibinfo {title} {Anisotropic-exchange magnets on a
  triangular lattice: Spin waves, accidental degeneracies, and dual spin
  liquids},\ }\href {https://doi.org/10.1103/PhysRevX.9.021017} {\bibfield
  {journal} {\bibinfo  {journal} {Phys. Rev. X}\ }\textbf {\bibinfo {volume}
  {9}},\ \bibinfo {pages} {021017} (\bibinfo {year} {2019})}\BibitemShut
  {NoStop}%
\bibitem [{\citenamefont {Li}\ \emph {et~al.}(2015)\citenamefont {Li},
  \citenamefont {Chen}, \citenamefont {Tong}, \citenamefont {Pi}, \citenamefont
  {Liu}, \citenamefont {Yang}, \citenamefont {Wang},\ and\ \citenamefont
  {Zhang}}]{Li167203}%
  \BibitemOpen
  \bibfield  {author} {\bibinfo {author} {\bibfnamefont {Y.}~\bibnamefont
  {Li}}, \bibinfo {author} {\bibfnamefont {G.}~\bibnamefont {Chen}}, \bibinfo
  {author} {\bibfnamefont {W.}~\bibnamefont {Tong}}, \bibinfo {author}
  {\bibfnamefont {L.}~\bibnamefont {Pi}}, \bibinfo {author} {\bibfnamefont
  {J.}~\bibnamefont {Liu}}, \bibinfo {author} {\bibfnamefont {Z.}~\bibnamefont
  {Yang}}, \bibinfo {author} {\bibfnamefont {X.}~\bibnamefont {Wang}},\ and\
  \bibinfo {author} {\bibfnamefont {Q.}~\bibnamefont {Zhang}},\ }\bibfield
  {title} {\bibinfo {title} {Rare-earth triangular lattice spin liquid: {A}
  single-crystal study of {YbMgGaO}$_4$},\ }\href
  {https://doi.org/10.1103/PhysRevLett.115.167203} {\bibfield  {journal}
  {\bibinfo  {journal} {Phys. Rev. Lett.}\ }\textbf {\bibinfo {volume} {115}},\
  \bibinfo {pages} {167203} (\bibinfo {year} {2015})}\BibitemShut {NoStop}%
\bibitem [{\citenamefont {Paddison}\ \emph {et~al.}(2017)\citenamefont
  {Paddison}, \citenamefont {Daum}, \citenamefont {Dun}, \citenamefont
  {Ehlers}, \citenamefont {Liu}, \citenamefont {Stone}, \citenamefont {Zhou},\
  and\ \citenamefont {Mourigal}}]{Paddison117}%
  \BibitemOpen
  \bibfield  {author} {\bibinfo {author} {\bibfnamefont {J.~A.}\ \bibnamefont
  {Paddison}}, \bibinfo {author} {\bibfnamefont {M.}~\bibnamefont {Daum}},
  \bibinfo {author} {\bibfnamefont {Z.}~\bibnamefont {Dun}}, \bibinfo {author}
  {\bibfnamefont {G.}~\bibnamefont {Ehlers}}, \bibinfo {author} {\bibfnamefont
  {Y.}~\bibnamefont {Liu}}, \bibinfo {author} {\bibfnamefont {M.}~\bibnamefont
  {Stone}}, \bibinfo {author} {\bibfnamefont {H.}~\bibnamefont {Zhou}},\ and\
  \bibinfo {author} {\bibfnamefont {M.}~\bibnamefont {Mourigal}},\ }\bibfield
  {title} {\bibinfo {title} {Continuous excitations of the triangular-lattice
  quantum spin liquid {YbMgGaO}$_4$},\ }\href
  {https://doi.org/10.1038/nphys3971} {\bibfield  {journal} {\bibinfo
  {journal} {Nat. Phys.}\ }\textbf {\bibinfo {volume} {13}},\ \bibinfo {pages}
  {117} (\bibinfo {year} {2017})}\BibitemShut {NoStop}%
\bibitem [{\citenamefont {Li}\ \emph {et~al.}(2017)\citenamefont {Li},
  \citenamefont {Adroja}, \citenamefont {Bewley}, \citenamefont {Voneshen},
  \citenamefont {Tsirlin}, \citenamefont {Gegenwart},\ and\ \citenamefont
  {Zhang}}]{Li107202}%
  \BibitemOpen
  \bibfield  {author} {\bibinfo {author} {\bibfnamefont {Y.}~\bibnamefont
  {Li}}, \bibinfo {author} {\bibfnamefont {D.}~\bibnamefont {Adroja}}, \bibinfo
  {author} {\bibfnamefont {R.~I.}\ \bibnamefont {Bewley}}, \bibinfo {author}
  {\bibfnamefont {D.}~\bibnamefont {Voneshen}}, \bibinfo {author}
  {\bibfnamefont {A.~A.}\ \bibnamefont {Tsirlin}}, \bibinfo {author}
  {\bibfnamefont {P.}~\bibnamefont {Gegenwart}},\ and\ \bibinfo {author}
  {\bibfnamefont {Q.}~\bibnamefont {Zhang}},\ }\bibfield  {title} {\bibinfo
  {title} {Crystalline electric-field randomness in the triangular lattice
  spin-liquid {YbMgGaO}$_4$},\ }\href
  {https://doi.org/10.1103/PhysRevLett.118.107202} {\bibfield  {journal}
  {\bibinfo  {journal} {Phys. Rev. Lett.}\ }\textbf {\bibinfo {volume} {118}},\
  \bibinfo {pages} {107202} (\bibinfo {year} {2017})}\BibitemShut {NoStop}%
\bibitem [{\citenamefont {Zhu}\ \emph {et~al.}(2017)\citenamefont {Zhu},
  \citenamefont {Maksimov}, \citenamefont {White},\ and\ \citenamefont
  {Chernyshev}}]{Zhu157201}%
  \BibitemOpen
  \bibfield  {author} {\bibinfo {author} {\bibfnamefont {Z.}~\bibnamefont
  {Zhu}}, \bibinfo {author} {\bibfnamefont {P.~A.}\ \bibnamefont {Maksimov}},
  \bibinfo {author} {\bibfnamefont {S.~R.}\ \bibnamefont {White}},\ and\
  \bibinfo {author} {\bibfnamefont {A.~L.}\ \bibnamefont {Chernyshev}},\
  }\bibfield  {title} {\bibinfo {title} {Disorder-induced mimicry of a spin
  liquid in {YbMgGaO}$_4$},\ }\href
  {https://doi.org/10.1103/PhysRevLett.119.157201} {\bibfield  {journal}
  {\bibinfo  {journal} {Phys. Rev. Lett.}\ }\textbf {\bibinfo {volume} {119}},\
  \bibinfo {pages} {157201} (\bibinfo {year} {2017})}\BibitemShut {NoStop}%
\bibitem [{\citenamefont {Kimchi}\ \emph {et~al.}(2018)\citenamefont {Kimchi},
  \citenamefont {Nahum},\ and\ \citenamefont {Senthil}}]{Kimchi031208}%
  \BibitemOpen
  \bibfield  {author} {\bibinfo {author} {\bibfnamefont {I.}~\bibnamefont
  {Kimchi}}, \bibinfo {author} {\bibfnamefont {A.}~\bibnamefont {Nahum}},\ and\
  \bibinfo {author} {\bibfnamefont {T.}~\bibnamefont {Senthil}},\ }\bibfield
  {title} {\bibinfo {title} {Valence bonds in random quantum magnets: Theory
  and application to {YbMgGaO}$_4$},\ }\href
  {https://doi.org/10.1103/PhysRevX.8.031028} {\bibfield  {journal} {\bibinfo
  {journal} {Phys. Rev. X}\ }\textbf {\bibinfo {volume} {8}},\ \bibinfo {pages}
  {031028} (\bibinfo {year} {2018})}\BibitemShut {NoStop}%
\bibitem [{\citenamefont {Ma}\ \emph {et~al.}(2018)\citenamefont {Ma},
  \citenamefont {Wang}, \citenamefont {Dong}, \citenamefont {Zhang},
  \citenamefont {Li}, \citenamefont {Zheng}, \citenamefont {Yu}, \citenamefont
  {Wang}, \citenamefont {Che}, \citenamefont {Ran}, \citenamefont {Bao},
  \citenamefont {Cai}, \citenamefont {\ifmmode~\check{C}\else
  \v{C}\fi{}erm\'ak}, \citenamefont {Schneidewind}, \citenamefont {Yano},
  \citenamefont {Gardner}, \citenamefont {Lu}, \citenamefont {Yu},
  \citenamefont {Liu}, \citenamefont {Li}, \citenamefont {Li},\ and\
  \citenamefont {Wen}}]{Ma087201}%
  \BibitemOpen
  \bibfield  {author} {\bibinfo {author} {\bibfnamefont {Z.}~\bibnamefont
  {Ma}}, \bibinfo {author} {\bibfnamefont {J.}~\bibnamefont {Wang}}, \bibinfo
  {author} {\bibfnamefont {Z.-Y.}\ \bibnamefont {Dong}}, \bibinfo {author}
  {\bibfnamefont {J.}~\bibnamefont {Zhang}}, \bibinfo {author} {\bibfnamefont
  {S.}~\bibnamefont {Li}}, \bibinfo {author} {\bibfnamefont {S.-H.}\
  \bibnamefont {Zheng}}, \bibinfo {author} {\bibfnamefont {Y.}~\bibnamefont
  {Yu}}, \bibinfo {author} {\bibfnamefont {W.}~\bibnamefont {Wang}}, \bibinfo
  {author} {\bibfnamefont {L.}~\bibnamefont {Che}}, \bibinfo {author}
  {\bibfnamefont {K.}~\bibnamefont {Ran}}, \bibinfo {author} {\bibfnamefont
  {S.}~\bibnamefont {Bao}}, \bibinfo {author} {\bibfnamefont {Z.}~\bibnamefont
  {Cai}}, \bibinfo {author} {\bibfnamefont {P.}~\bibnamefont
  {\ifmmode~\check{C}\else \v{C}\fi{}erm\'ak}}, \bibinfo {author}
  {\bibfnamefont {A.}~\bibnamefont {Schneidewind}}, \bibinfo {author}
  {\bibfnamefont {S.}~\bibnamefont {Yano}}, \bibinfo {author} {\bibfnamefont
  {J.~S.}\ \bibnamefont {Gardner}}, \bibinfo {author} {\bibfnamefont
  {X.}~\bibnamefont {Lu}}, \bibinfo {author} {\bibfnamefont {S.-L.}\
  \bibnamefont {Yu}}, \bibinfo {author} {\bibfnamefont {J.-M.}\ \bibnamefont
  {Liu}}, \bibinfo {author} {\bibfnamefont {S.}~\bibnamefont {Li}}, \bibinfo
  {author} {\bibfnamefont {J.-X.}\ \bibnamefont {Li}},\ and\ \bibinfo {author}
  {\bibfnamefont {J.}~\bibnamefont {Wen}},\ }\bibfield  {title} {\bibinfo
  {title} {Spin-glass ground state in a triangular-lattice compound
  {YbZnGaO}$_4$},\ }\href {https://doi.org/10.1103/PhysRevLett.120.087201}
  {\bibfield  {journal} {\bibinfo  {journal} {Phys. Rev. Lett.}\ }\textbf
  {\bibinfo {volume} {120}},\ \bibinfo {pages} {087201} (\bibinfo {year}
  {2018})}\BibitemShut {NoStop}%
\bibitem [{\citenamefont {Ma}\ \emph {et~al.}(2021)\citenamefont {Ma},
  \citenamefont {Dong}, \citenamefont {Wang}, \citenamefont {Zheng},
  \citenamefont {Ran}, \citenamefont {Bao}, \citenamefont {Cai}, \citenamefont
  {Shangguan}, \citenamefont {Wang}, \citenamefont {Boehm}, \citenamefont
  {Steffens}, \citenamefont {Regnault}, \citenamefont {Wang}, \citenamefont
  {Su}, \citenamefont {Yu}, \citenamefont {Liu}, \citenamefont {Li},\ and\
  \citenamefont {Wen}}]{Ma224433}%
  \BibitemOpen
  \bibfield  {author} {\bibinfo {author} {\bibfnamefont {Z.}~\bibnamefont
  {Ma}}, \bibinfo {author} {\bibfnamefont {Z.-Y.}\ \bibnamefont {Dong}},
  \bibinfo {author} {\bibfnamefont {J.}~\bibnamefont {Wang}}, \bibinfo {author}
  {\bibfnamefont {S.}~\bibnamefont {Zheng}}, \bibinfo {author} {\bibfnamefont
  {K.}~\bibnamefont {Ran}}, \bibinfo {author} {\bibfnamefont {S.}~\bibnamefont
  {Bao}}, \bibinfo {author} {\bibfnamefont {Z.}~\bibnamefont {Cai}}, \bibinfo
  {author} {\bibfnamefont {Y.}~\bibnamefont {Shangguan}}, \bibinfo {author}
  {\bibfnamefont {W.}~\bibnamefont {Wang}}, \bibinfo {author} {\bibfnamefont
  {M.}~\bibnamefont {Boehm}}, \bibinfo {author} {\bibfnamefont
  {P.}~\bibnamefont {Steffens}}, \bibinfo {author} {\bibfnamefont {L.-P.}\
  \bibnamefont {Regnault}}, \bibinfo {author} {\bibfnamefont {X.}~\bibnamefont
  {Wang}}, \bibinfo {author} {\bibfnamefont {Y.}~\bibnamefont {Su}}, \bibinfo
  {author} {\bibfnamefont {S.-L.}\ \bibnamefont {Yu}}, \bibinfo {author}
  {\bibfnamefont {J.-M.}\ \bibnamefont {Liu}}, \bibinfo {author} {\bibfnamefont
  {J.-X.}\ \bibnamefont {Li}},\ and\ \bibinfo {author} {\bibfnamefont
  {J.}~\bibnamefont {Wen}},\ }\bibfield  {title} {\bibinfo {title}
  {Disorder-induced broadening of the spin waves in the triangular-lattice
  quantum spin liquid candidate {YbZnGaO}$_4$},\ }\href
  {https://doi.org/10.1103/PhysRevB.104.224433} {\bibfield  {journal} {\bibinfo
   {journal} {Phys. Rev. B}\ }\textbf {\bibinfo {volume} {104}},\ \bibinfo
  {pages} {224433} (\bibinfo {year} {2021})}\BibitemShut {NoStop}%
\bibitem [{\citenamefont {Rao}\ \emph {et~al.}(2021)\citenamefont {Rao},
  \citenamefont {Hussain}, \citenamefont {Huang}, \citenamefont {Chu},
  \citenamefont {Li}, \citenamefont {Zhao}, \citenamefont {Dun}, \citenamefont
  {Choi}, \citenamefont {Asaba}, \citenamefont {Chen}, \citenamefont {Li},
  \citenamefont {Yue}, \citenamefont {Wang}, \citenamefont {Cheng},
  \citenamefont {Gao}, \citenamefont {Shen}, \citenamefont {Zhao},
  \citenamefont {Chen}, \citenamefont {Zhou},\ and\ \citenamefont
  {Sun}}]{Rao2021}%
  \BibitemOpen
  \bibfield  {author} {\bibinfo {author} {\bibfnamefont {X.}~\bibnamefont
  {Rao}}, \bibinfo {author} {\bibfnamefont {G.}~\bibnamefont {Hussain}},
  \bibinfo {author} {\bibfnamefont {Q.}~\bibnamefont {Huang}}, \bibinfo
  {author} {\bibfnamefont {W.~J.}\ \bibnamefont {Chu}}, \bibinfo {author}
  {\bibfnamefont {N.}~\bibnamefont {Li}}, \bibinfo {author} {\bibfnamefont
  {X.}~\bibnamefont {Zhao}}, \bibinfo {author} {\bibfnamefont {Z.}~\bibnamefont
  {Dun}}, \bibinfo {author} {\bibfnamefont {E.~S.}\ \bibnamefont {Choi}},
  \bibinfo {author} {\bibfnamefont {T.}~\bibnamefont {Asaba}}, \bibinfo
  {author} {\bibfnamefont {L.}~\bibnamefont {Chen}}, \bibinfo {author}
  {\bibfnamefont {L.}~\bibnamefont {Li}}, \bibinfo {author} {\bibfnamefont
  {X.~Y.}\ \bibnamefont {Yue}}, \bibinfo {author} {\bibfnamefont {N.~N.}\
  \bibnamefont {Wang}}, \bibinfo {author} {\bibfnamefont {J.-G.}\ \bibnamefont
  {Cheng}}, \bibinfo {author} {\bibfnamefont {Y.~H.}\ \bibnamefont {Gao}},
  \bibinfo {author} {\bibfnamefont {Y.}~\bibnamefont {Shen}}, \bibinfo {author}
  {\bibfnamefont {J.}~\bibnamefont {Zhao}}, \bibinfo {author} {\bibfnamefont
  {G.}~\bibnamefont {Chen}}, \bibinfo {author} {\bibfnamefont {H.~D.}\
  \bibnamefont {Zhou}},\ and\ \bibinfo {author} {\bibfnamefont {X.~F.}\
  \bibnamefont {Sun}},\ }\bibfield  {title} {\bibinfo {title} {Survival of
  itinerant excitations and quantum spin state transitions in {YbMgGaO}$_4$
  with chemical disorder},\ }\href {https://doi.org/10.1038/s41467-021-25247-6}
  {\bibfield  {journal} {\bibinfo  {journal} {Nat. Commun.}\ }\textbf {\bibinfo
  {volume} {12}},\ \bibinfo {pages} {4949} (\bibinfo {year}
  {2021})}\BibitemShut {NoStop}%
\bibitem [{\citenamefont {Li}\ \emph {et~al.}(2019)\citenamefont {Li},
  \citenamefont {Bachus}, \citenamefont {Liu}, \citenamefont {Radelytskyi},
  \citenamefont {Bertin}, \citenamefont {Schneidewind}, \citenamefont {Tokiwa},
  \citenamefont {Tsirlin},\ and\ \citenamefont {Gegenwart}}]{Li137201}%
  \BibitemOpen
  \bibfield  {author} {\bibinfo {author} {\bibfnamefont {Y.}~\bibnamefont
  {Li}}, \bibinfo {author} {\bibfnamefont {S.}~\bibnamefont {Bachus}}, \bibinfo
  {author} {\bibfnamefont {B.}~\bibnamefont {Liu}}, \bibinfo {author}
  {\bibfnamefont {I.}~\bibnamefont {Radelytskyi}}, \bibinfo {author}
  {\bibfnamefont {A.}~\bibnamefont {Bertin}}, \bibinfo {author} {\bibfnamefont
  {A.}~\bibnamefont {Schneidewind}}, \bibinfo {author} {\bibfnamefont
  {Y.}~\bibnamefont {Tokiwa}}, \bibinfo {author} {\bibfnamefont {A.~A.}\
  \bibnamefont {Tsirlin}},\ and\ \bibinfo {author} {\bibfnamefont
  {P.}~\bibnamefont {Gegenwart}},\ }\bibfield  {title} {\bibinfo {title}
  {Rearrangement of uncorrelated valence bonds evidenced by low-energy spin
  excitations in {YbMgGaO}$_4$},\ }\href
  {https://doi.org/10.1103/PhysRevLett.122.137201} {\bibfield  {journal}
  {\bibinfo  {journal} {Phys. Rev. Lett.}\ }\textbf {\bibinfo {volume} {122}},\
  \bibinfo {pages} {137201} (\bibinfo {year} {2019})}\BibitemShut {NoStop}%
\bibitem [{\citenamefont {Bag}\ \emph {et~al.}(2024)\citenamefont {Bag},
  \citenamefont {Xu}, \citenamefont {Sherman}, \citenamefont {Yadav},
  \citenamefont {Kolesnikov}, \citenamefont {Podlesnyak}, \citenamefont {Choi},
  \citenamefont {da~Silva}, \citenamefont {Moore},\ and\ \citenamefont
  {Haravifard}}]{Bag266703}%
  \BibitemOpen
  \bibfield  {author} {\bibinfo {author} {\bibfnamefont {R.}~\bibnamefont
  {Bag}}, \bibinfo {author} {\bibfnamefont {S.}~\bibnamefont {Xu}}, \bibinfo
  {author} {\bibfnamefont {N.~E.}\ \bibnamefont {Sherman}}, \bibinfo {author}
  {\bibfnamefont {L.}~\bibnamefont {Yadav}}, \bibinfo {author} {\bibfnamefont
  {A.~I.}\ \bibnamefont {Kolesnikov}}, \bibinfo {author} {\bibfnamefont
  {A.~A.}\ \bibnamefont {Podlesnyak}}, \bibinfo {author} {\bibfnamefont
  {E.~S.}\ \bibnamefont {Choi}}, \bibinfo {author} {\bibfnamefont
  {I.}~\bibnamefont {da~Silva}}, \bibinfo {author} {\bibfnamefont {J.~E.}\
  \bibnamefont {Moore}},\ and\ \bibinfo {author} {\bibfnamefont
  {S.}~\bibnamefont {Haravifard}},\ }\bibfield  {title} {\bibinfo {title}
  {Evidence of {D}irac quantum spin liquid in {YbZn}$_2${GaO}$_5$},\ }\href
  {https://doi.org/10.1103/PhysRevLett.133.266703} {\bibfield  {journal}
  {\bibinfo  {journal} {Phys. Rev. Lett.}\ }\textbf {\bibinfo {volume} {133}},\
  \bibinfo {pages} {266703} (\bibinfo {year} {2024})}\BibitemShut {NoStop}%
\bibitem [{\citenamefont {Hu}\ \emph {et~al.}(2019)\citenamefont {Hu},
  \citenamefont {Zhu}, \citenamefont {Eggert},\ and\ \citenamefont
  {He}}]{Hu207203}%
  \BibitemOpen
  \bibfield  {author} {\bibinfo {author} {\bibfnamefont {S.}~\bibnamefont
  {Hu}}, \bibinfo {author} {\bibfnamefont {W.}~\bibnamefont {Zhu}}, \bibinfo
  {author} {\bibfnamefont {S.}~\bibnamefont {Eggert}},\ and\ \bibinfo {author}
  {\bibfnamefont {Y.-C.}\ \bibnamefont {He}},\ }\bibfield  {title} {\bibinfo
  {title} {Dirac spin liquid on the spin-$1/2$ triangular {H}eisenberg
  antiferromagnet},\ }\href {https://doi.org/10.1103/PhysRevLett.123.207203}
  {\bibfield  {journal} {\bibinfo  {journal} {Phys. Rev. Lett.}\ }\textbf
  {\bibinfo {volume} {123}},\ \bibinfo {pages} {207203} (\bibinfo {year}
  {2019})}\BibitemShut {NoStop}%
\bibitem [{\citenamefont {Wu}\ \emph {et~al.}(2025)\citenamefont {Wu},
  \citenamefont {Pratt}, \citenamefont {Huddart}, \citenamefont {Chatterjee},
  \citenamefont {Goddard}, \citenamefont {Singleton}, \citenamefont
  {Prabhakaran},\ and\ \citenamefont {Blundell}}]{wu2025spin}%
  \BibitemOpen
  \bibfield  {author} {\bibinfo {author} {\bibfnamefont {H.~C.}\ \bibnamefont
  {Wu}}, \bibinfo {author} {\bibfnamefont {F.~L.}\ \bibnamefont {Pratt}},
  \bibinfo {author} {\bibfnamefont {B.~M.}\ \bibnamefont {Huddart}}, \bibinfo
  {author} {\bibfnamefont {D.}~\bibnamefont {Chatterjee}}, \bibinfo {author}
  {\bibfnamefont {P.~A.}\ \bibnamefont {Goddard}}, \bibinfo {author}
  {\bibfnamefont {J.}~\bibnamefont {Singleton}}, \bibinfo {author}
  {\bibfnamefont {D.}~\bibnamefont {Prabhakaran}},\ and\ \bibinfo {author}
  {\bibfnamefont {S.~J.}\ \bibnamefont {Blundell}},\ }\bibfield  {title}
  {\bibinfo {title} {Spin dynamics in the {D}irac ${U}(1)$ spin liquid
  {YbZn}$_2${GaO}$_5$},\ }\href {https://doi.org/10.1103/l93v-f576} {\bibfield
  {journal} {\bibinfo  {journal} {Phys. Rev. Lett.}\ }\textbf {\bibinfo
  {volume} {135}},\ \bibinfo {pages} {046704} (\bibinfo {year}
  {2025})}\BibitemShut {NoStop}%
\bibitem [{\citenamefont {Zhao}\ \emph {et~al.}(2026)\citenamefont {Zhao},
  \citenamefont {Chen}, \citenamefont {Stone}, \citenamefont {Zhang},
  \citenamefont {Sarkis}, \citenamefont {Koohpayeh},\ and\ \citenamefont
  {Broholm}}]{zhao2025quenched}%
  \BibitemOpen
  \bibfield  {author} {\bibinfo {author} {\bibfnamefont {L.}~\bibnamefont
  {Zhao}}, \bibinfo {author} {\bibfnamefont {T.}~\bibnamefont {Chen}}, \bibinfo
  {author} {\bibfnamefont {M.~B.}\ \bibnamefont {Stone}}, \bibinfo {author}
  {\bibfnamefont {Q.}~\bibnamefont {Zhang}}, \bibinfo {author} {\bibfnamefont
  {C.~L.}\ \bibnamefont {Sarkis}}, \bibinfo {author} {\bibfnamefont {S.~M.}\
  \bibnamefont {Koohpayeh}},\ and\ \bibinfo {author} {\bibfnamefont {C.~L.}\
  \bibnamefont {Broholm}},\ }\bibfield  {title} {\bibinfo {title} {Quenched
  disorder in the triangular lattice antiferromagnet {YbZn}$_2${GaO}$_5$},\
  }\href {https://doi.org/10.1103/xn2m-1jb5} {\bibfield  {journal} {\bibinfo
  {journal} {Phys. Rev. B}\ }\textbf {\bibinfo {volume} {113}},\ \bibinfo
  {pages} {014437} (\bibinfo {year} {2026})}\BibitemShut {NoStop}%
\bibitem [{SM()}]{SM}%
  \BibitemOpen
  \href@noop {} {}\bibinfo {note} {See Supplemental Material at
  \url{http://link.aps.org/supplemental/10.1103/n1ly-lsck} for more details
  about NMR and $\mu$SR data analysis, which also includes
  Refs.~[18,28,32,41].}\BibitemShut {Stop}%
\bibitem [{\citenamefont {Slichter}(1990)}]{slichter2013principles}%
  \BibitemOpen
  \bibfield  {author} {\bibinfo {author} {\bibfnamefont {C.~P.}\ \bibnamefont
  {Slichter}},\ }\href
  {https://link.springer.com/book/10.1007/978-3-662-09441-9} {\emph {\bibinfo
  {title} {Principles of magnetic resonance}}},\ \bibinfo {edition} {3rd}\ ed.\
  (\bibinfo  {publisher} {Springer},\ \bibinfo {address} {Berlin},\ \bibinfo
  {year} {1990})\BibitemShut {NoStop}%
\bibitem [{\citenamefont {Suter}\ \emph {et~al.}(1998)\citenamefont {Suter},
  \citenamefont {Mali}, \citenamefont {Roos},\ and\ \citenamefont
  {Brinkmann}}]{suter1998mixed}%
  \BibitemOpen
  \bibfield  {author} {\bibinfo {author} {\bibfnamefont {A.}~\bibnamefont
  {Suter}}, \bibinfo {author} {\bibfnamefont {M.}~\bibnamefont {Mali}},
  \bibinfo {author} {\bibfnamefont {J.}~\bibnamefont {Roos}},\ and\ \bibinfo
  {author} {\bibfnamefont {D.}~\bibnamefont {Brinkmann}},\ }\bibfield  {title}
  {\bibinfo {title} {Mixed magnetic and quadrupolar relaxation in the presence
  of a dominant static {Z}eeman {H}amiltonian},\ }\href
  {https://doi.org/10.1088/0953-8984/10/26/022} {\bibfield  {journal} {\bibinfo
   {journal} {J. Phys.: Condens. Matter}\ }\textbf {\bibinfo {volume} {10}},\
  \bibinfo {pages} {5977} (\bibinfo {year} {1998})}\BibitemShut {NoStop}%
\bibitem [{\citenamefont {Ranjith}\ \emph {et~al.}(2019)\citenamefont
  {Ranjith}, \citenamefont {Dmytriieva}, \citenamefont {Khim}, \citenamefont
  {Sichelschmidt}, \citenamefont {Luther}, \citenamefont {Ehlers},
  \citenamefont {Yasuoka}, \citenamefont {Wosnitza}, \citenamefont {Tsirlin},
  \citenamefont {K{\"u}hne} \emph {et~al.}}]{ranjith2019field}%
  \BibitemOpen
  \bibfield  {author} {\bibinfo {author} {\bibfnamefont {K.}~\bibnamefont
  {Ranjith}}, \bibinfo {author} {\bibfnamefont {D.}~\bibnamefont {Dmytriieva}},
  \bibinfo {author} {\bibfnamefont {S.}~\bibnamefont {Khim}}, \bibinfo {author}
  {\bibfnamefont {J.}~\bibnamefont {Sichelschmidt}}, \bibinfo {author}
  {\bibfnamefont {S.}~\bibnamefont {Luther}}, \bibinfo {author} {\bibfnamefont
  {D.}~\bibnamefont {Ehlers}}, \bibinfo {author} {\bibfnamefont
  {H.}~\bibnamefont {Yasuoka}}, \bibinfo {author} {\bibfnamefont
  {J.}~\bibnamefont {Wosnitza}}, \bibinfo {author} {\bibfnamefont {A.~A.}\
  \bibnamefont {Tsirlin}}, \bibinfo {author} {\bibfnamefont {H.}~\bibnamefont
  {K{\"u}hne}}, \emph {et~al.},\ }\bibfield  {title} {\bibinfo {title}
  {Field-induced instability of the quantum spin liquid ground state in the j
  eff= 1 2 triangular-lattice compound naybo 2},\ }\href@noop {} {\bibfield
  {journal} {\bibinfo  {journal} {Physical Review B}\ }\textbf {\bibinfo
  {volume} {99}},\ \bibinfo {pages} {180401} (\bibinfo {year}
  {2019})}\BibitemShut {NoStop}%
\bibitem [{\citenamefont {Amato}\ and\ \citenamefont
  {Morenzoni}(2024)}]{Amato2024}%
  \BibitemOpen
  \bibfield  {author} {\bibinfo {author} {\bibfnamefont {A.}~\bibnamefont
  {Amato}}\ and\ \bibinfo {author} {\bibfnamefont {E.}~\bibnamefont
  {Morenzoni}},\ }\href
  {https://doi.org/https://doi.org/10.1007/978-3-031-44959-8} {\emph {\bibinfo
  {title} {Introduction to Muon Spin Spectroscopy: Applications to Solid State
  and Material Sciences}}}\ (\bibinfo  {publisher} {Springer},\ \bibinfo
  {address} {Cham},\ \bibinfo {year} {2024})\BibitemShut {NoStop}%
\bibitem [{\citenamefont {Mendels}\ \emph {et~al.}(2007)\citenamefont
  {Mendels}, \citenamefont {Bert}, \citenamefont {de~Vries}, \citenamefont
  {Olariu}, \citenamefont {Harrison}, \citenamefont {Duc}, \citenamefont
  {Trombe}, \citenamefont {Lord}, \citenamefont {Amato},\ and\ \citenamefont
  {Baines}}]{Mendels077204}%
  \BibitemOpen
  \bibfield  {author} {\bibinfo {author} {\bibfnamefont {P.}~\bibnamefont
  {Mendels}}, \bibinfo {author} {\bibfnamefont {F.}~\bibnamefont {Bert}},
  \bibinfo {author} {\bibfnamefont {M.~A.}\ \bibnamefont {de~Vries}}, \bibinfo
  {author} {\bibfnamefont {A.}~\bibnamefont {Olariu}}, \bibinfo {author}
  {\bibfnamefont {A.}~\bibnamefont {Harrison}}, \bibinfo {author}
  {\bibfnamefont {F.}~\bibnamefont {Duc}}, \bibinfo {author} {\bibfnamefont
  {J.~C.}\ \bibnamefont {Trombe}}, \bibinfo {author} {\bibfnamefont {J.~S.}\
  \bibnamefont {Lord}}, \bibinfo {author} {\bibfnamefont {A.}~\bibnamefont
  {Amato}},\ and\ \bibinfo {author} {\bibfnamefont {C.}~\bibnamefont
  {Baines}},\ }\bibfield  {title} {\bibinfo {title} {Quantum magnetism in the
  paratacamite family: {T}owards an ideal kagom\'e lattice},\ }\href
  {https://doi.org/10.1103/PhysRevLett.98.077204} {\bibfield  {journal}
  {\bibinfo  {journal} {Phys. Rev. Lett.}\ }\textbf {\bibinfo {volume} {98}},\
  \bibinfo {pages} {077204} (\bibinfo {year} {2007})}\BibitemShut {NoStop}%
\bibitem [{\citenamefont {Uemura}\ \emph {et~al.}(1985)\citenamefont {Uemura},
  \citenamefont {Yamazaki}, \citenamefont {Harshman}, \citenamefont {Senba},\
  and\ \citenamefont {Ansaldo}}]{Uemura546}%
  \BibitemOpen
  \bibfield  {author} {\bibinfo {author} {\bibfnamefont {Y.~J.}\ \bibnamefont
  {Uemura}}, \bibinfo {author} {\bibfnamefont {T.}~\bibnamefont {Yamazaki}},
  \bibinfo {author} {\bibfnamefont {D.~R.}\ \bibnamefont {Harshman}}, \bibinfo
  {author} {\bibfnamefont {M.}~\bibnamefont {Senba}},\ and\ \bibinfo {author}
  {\bibfnamefont {E.~J.}\ \bibnamefont {Ansaldo}},\ }\bibfield  {title}
  {\bibinfo {title} {Muon-spin relaxation in {AuFe} and {CuMn} spin glasses},\
  }\href {https://doi.org/10.1103/PhysRevB.31.546} {\bibfield  {journal}
  {\bibinfo  {journal} {Phys. Rev. B}\ }\textbf {\bibinfo {volume} {31}},\
  \bibinfo {pages} {546} (\bibinfo {year} {1985})}\BibitemShut {NoStop}%
\bibitem [{\citenamefont {Uemura}\ \emph {et~al.}(1994)\citenamefont {Uemura},
  \citenamefont {Keren}, \citenamefont {Kojima}, \citenamefont {Le},
  \citenamefont {Luke}, \citenamefont {Wu}, \citenamefont {Ajiro},
  \citenamefont {Asano}, \citenamefont {Kuriyama}, \citenamefont {Mekata},
  \citenamefont {Kikuchi},\ and\ \citenamefont {Kakurai}}]{Uemura3306}%
  \BibitemOpen
  \bibfield  {author} {\bibinfo {author} {\bibfnamefont {Y.~J.}\ \bibnamefont
  {Uemura}}, \bibinfo {author} {\bibfnamefont {A.}~\bibnamefont {Keren}},
  \bibinfo {author} {\bibfnamefont {K.}~\bibnamefont {Kojima}}, \bibinfo
  {author} {\bibfnamefont {L.~P.}\ \bibnamefont {Le}}, \bibinfo {author}
  {\bibfnamefont {G.~M.}\ \bibnamefont {Luke}}, \bibinfo {author}
  {\bibfnamefont {W.~D.}\ \bibnamefont {Wu}}, \bibinfo {author} {\bibfnamefont
  {Y.}~\bibnamefont {Ajiro}}, \bibinfo {author} {\bibfnamefont
  {T.}~\bibnamefont {Asano}}, \bibinfo {author} {\bibfnamefont
  {Y.}~\bibnamefont {Kuriyama}}, \bibinfo {author} {\bibfnamefont
  {M.}~\bibnamefont {Mekata}}, \bibinfo {author} {\bibfnamefont
  {H.}~\bibnamefont {Kikuchi}},\ and\ \bibinfo {author} {\bibfnamefont
  {K.}~\bibnamefont {Kakurai}},\ }\bibfield  {title} {\bibinfo {title} {Spin
  fluctuations in frustrated kagom\'e lattice system {$\rm SrCr_8Ga_4O_{19}$}
  studied by muon spin relaxation},\ }\href
  {https://doi.org/10.1103/PhysRevLett.73.3306} {\bibfield  {journal} {\bibinfo
   {journal} {Phys. Rev. Lett.}\ }\textbf {\bibinfo {volume} {73}},\ \bibinfo
  {pages} {3306} (\bibinfo {year} {1994})}\BibitemShut {NoStop}%
\bibitem [{\citenamefont {Li}\ \emph {et~al.}(2016)\citenamefont {Li},
  \citenamefont {Adroja}, \citenamefont {Biswas}, \citenamefont {Baker},
  \citenamefont {Zhang}, \citenamefont {Liu}, \citenamefont {Tsirlin},
  \citenamefont {Gegenwart},\ and\ \citenamefont {Zhang}}]{Li097201}%
  \BibitemOpen
  \bibfield  {author} {\bibinfo {author} {\bibfnamefont {Y.}~\bibnamefont
  {Li}}, \bibinfo {author} {\bibfnamefont {D.}~\bibnamefont {Adroja}}, \bibinfo
  {author} {\bibfnamefont {P.~K.}\ \bibnamefont {Biswas}}, \bibinfo {author}
  {\bibfnamefont {P.~J.}\ \bibnamefont {Baker}}, \bibinfo {author}
  {\bibfnamefont {Q.}~\bibnamefont {Zhang}}, \bibinfo {author} {\bibfnamefont
  {J.}~\bibnamefont {Liu}}, \bibinfo {author} {\bibfnamefont {A.~A.}\
  \bibnamefont {Tsirlin}}, \bibinfo {author} {\bibfnamefont {P.}~\bibnamefont
  {Gegenwart}},\ and\ \bibinfo {author} {\bibfnamefont {Q.}~\bibnamefont
  {Zhang}},\ }\bibfield  {title} {\bibinfo {title} {Muon spin relaxation
  evidence for the ${U}(1)$ quantum spin-liquid ground state in the triangular
  antiferromagnet {YbMgGaO}$_4$},\ }\href
  {https://doi.org/10.1103/PhysRevLett.117.097201} {\bibfield  {journal}
  {\bibinfo  {journal} {Phys. Rev. Lett.}\ }\textbf {\bibinfo {volume} {117}},\
  \bibinfo {pages} {097201} (\bibinfo {year} {2016})}\BibitemShut {NoStop}%
\bibitem [{\citenamefont {Redfield}(1965)}]{Redfield1965}%
  \BibitemOpen
  \bibfield  {author} {\bibinfo {author} {\bibfnamefont {A.~G.}\ \bibnamefont
  {Redfield}},\ }\bibfield  {title} {\bibinfo {title} {The theory of relaxation
  processes},\ }in\ \href
  {https://doi.org/https://doi.org/10.1016/B978-1-4832-3114-3.50007-6} {\emph
  {\bibinfo {booktitle} {Adv. Magn. Reson.}}},\ Vol.~\bibinfo {volume} {1},\
  \bibinfo {editor} {edited by\ \bibinfo {editor} {\bibfnamefont {J.~S.}\
  \bibnamefont {Waugh}}}\ (\bibinfo  {publisher} {Academic Press},\ \bibinfo
  {address} {New York},\ \bibinfo {year} {1965})\ pp.\ \bibinfo {pages}
  {1--32}\BibitemShut {NoStop}%
\bibitem [{\citenamefont {F\aa{}k}\ \emph {et~al.}(2012)\citenamefont
  {F\aa{}k}, \citenamefont {Kermarrec}, \citenamefont {Messio}, \citenamefont
  {Bernu}, \citenamefont {Lhuillier}, \citenamefont {Bert}, \citenamefont
  {Mendels}, \citenamefont {Koteswararao}, \citenamefont {Bouquet},
  \citenamefont {Ollivier}, \citenamefont {Hillier}, \citenamefont {Amato},
  \citenamefont {Colman},\ and\ \citenamefont {Wills}}]{Faak037208}%
  \BibitemOpen
  \bibfield  {author} {\bibinfo {author} {\bibfnamefont {B.}~\bibnamefont
  {F\aa{}k}}, \bibinfo {author} {\bibfnamefont {E.}~\bibnamefont {Kermarrec}},
  \bibinfo {author} {\bibfnamefont {L.}~\bibnamefont {Messio}}, \bibinfo
  {author} {\bibfnamefont {B.}~\bibnamefont {Bernu}}, \bibinfo {author}
  {\bibfnamefont {C.}~\bibnamefont {Lhuillier}}, \bibinfo {author}
  {\bibfnamefont {F.}~\bibnamefont {Bert}}, \bibinfo {author} {\bibfnamefont
  {P.}~\bibnamefont {Mendels}}, \bibinfo {author} {\bibfnamefont
  {B.}~\bibnamefont {Koteswararao}}, \bibinfo {author} {\bibfnamefont
  {F.}~\bibnamefont {Bouquet}}, \bibinfo {author} {\bibfnamefont
  {J.}~\bibnamefont {Ollivier}}, \bibinfo {author} {\bibfnamefont {A.~D.}\
  \bibnamefont {Hillier}}, \bibinfo {author} {\bibfnamefont {A.}~\bibnamefont
  {Amato}}, \bibinfo {author} {\bibfnamefont {R.~H.}\ \bibnamefont {Colman}},\
  and\ \bibinfo {author} {\bibfnamefont {A.~S.}\ \bibnamefont {Wills}},\
  }\bibfield  {title} {\bibinfo {title} {Kapellasite: A kagome quantum spin
  liquid with competing interactions},\ }\href
  {https://doi.org/10.1103/PhysRevLett.109.037208} {\bibfield  {journal}
  {\bibinfo  {journal} {Phys. Rev. Lett.}\ }\textbf {\bibinfo {volume} {109}},\
  \bibinfo {pages} {037208} (\bibinfo {year} {2012})}\BibitemShut {NoStop}%
\bibitem [{\citenamefont {Quilliam}\ \emph {et~al.}(2016)\citenamefont
  {Quilliam}, \citenamefont {Bert}, \citenamefont {Manseau}, \citenamefont
  {Darie}, \citenamefont {Guillot-Deudon}, \citenamefont {Payen}, \citenamefont
  {Baines}, \citenamefont {Amato},\ and\ \citenamefont
  {Mendels}}]{Quilliam214432}%
  \BibitemOpen
  \bibfield  {author} {\bibinfo {author} {\bibfnamefont {J.~A.}\ \bibnamefont
  {Quilliam}}, \bibinfo {author} {\bibfnamefont {F.}~\bibnamefont {Bert}},
  \bibinfo {author} {\bibfnamefont {A.}~\bibnamefont {Manseau}}, \bibinfo
  {author} {\bibfnamefont {C.}~\bibnamefont {Darie}}, \bibinfo {author}
  {\bibfnamefont {C.}~\bibnamefont {Guillot-Deudon}}, \bibinfo {author}
  {\bibfnamefont {C.}~\bibnamefont {Payen}}, \bibinfo {author} {\bibfnamefont
  {C.}~\bibnamefont {Baines}}, \bibinfo {author} {\bibfnamefont
  {A.}~\bibnamefont {Amato}},\ and\ \bibinfo {author} {\bibfnamefont
  {P.}~\bibnamefont {Mendels}},\ }\bibfield  {title} {\bibinfo {title} {Gapless
  quantum spin liquid ground state in the spin-1 antiferromagnet
  {6HB}-{Ba}$_3${NiSb}$_2${O}$_9$},\ }\href
  {https://doi.org/10.1103/PhysRevB.93.214432} {\bibfield  {journal} {\bibinfo
  {journal} {Phys. Rev. B}\ }\textbf {\bibinfo {volume} {93}},\ \bibinfo
  {pages} {214432} (\bibinfo {year} {2016})}\BibitemShut {NoStop}%
\bibitem [{\citenamefont {Keren}\ \emph {et~al.}(2001)\citenamefont {Keren},
  \citenamefont {Bazalitsky}, \citenamefont {Campbell},\ and\ \citenamefont
  {Lord}}]{Keren054403}%
  \BibitemOpen
  \bibfield  {author} {\bibinfo {author} {\bibfnamefont {A.}~\bibnamefont
  {Keren}}, \bibinfo {author} {\bibfnamefont {G.}~\bibnamefont {Bazalitsky}},
  \bibinfo {author} {\bibfnamefont {I.}~\bibnamefont {Campbell}},\ and\
  \bibinfo {author} {\bibfnamefont {J.~S.}\ \bibnamefont {Lord}},\ }\bibfield
  {title} {\bibinfo {title} {Probing exotic spin correlations by muon spin
  depolarization measurements with applications to spin glass dynamics},\
  }\href {https://doi.org/10.1103/PhysRevB.64.054403} {\bibfield  {journal}
  {\bibinfo  {journal} {Phys. Rev. B}\ }\textbf {\bibinfo {volume} {64}},\
  \bibinfo {pages} {054403} (\bibinfo {year} {2001})}\BibitemShut {NoStop}%
\bibitem [{\citenamefont {Kermarrec}\ \emph {et~al.}(2011)\citenamefont
  {Kermarrec}, \citenamefont {Mendels}, \citenamefont {Bert}, \citenamefont
  {Colman}, \citenamefont {Wills}, \citenamefont {Strobel}, \citenamefont
  {Bonville}, \citenamefont {Hillier},\ and\ \citenamefont
  {Amato}}]{Kermarrec100401}%
  \BibitemOpen
  \bibfield  {author} {\bibinfo {author} {\bibfnamefont {E.}~\bibnamefont
  {Kermarrec}}, \bibinfo {author} {\bibfnamefont {P.}~\bibnamefont {Mendels}},
  \bibinfo {author} {\bibfnamefont {F.}~\bibnamefont {Bert}}, \bibinfo {author}
  {\bibfnamefont {R.~H.}\ \bibnamefont {Colman}}, \bibinfo {author}
  {\bibfnamefont {A.~S.}\ \bibnamefont {Wills}}, \bibinfo {author}
  {\bibfnamefont {P.}~\bibnamefont {Strobel}}, \bibinfo {author} {\bibfnamefont
  {P.}~\bibnamefont {Bonville}}, \bibinfo {author} {\bibfnamefont
  {A.}~\bibnamefont {Hillier}},\ and\ \bibinfo {author} {\bibfnamefont
  {A.}~\bibnamefont {Amato}},\ }\bibfield  {title} {\bibinfo {title}
  {Spin-liquid ground state in the frustrated kagome antiferromagnet
  {MgCu}$_3${(OH)}$_6${Cl}$_2$},\ }\href
  {https://doi.org/10.1103/PhysRevB.84.100401} {\bibfield  {journal} {\bibinfo
  {journal} {Phys. Rev. B}\ }\textbf {\bibinfo {volume} {84}},\ \bibinfo
  {pages} {100401} (\bibinfo {year} {2011})}\BibitemShut {NoStop}%
\bibitem [{\citenamefont {Keren}\ \emph {et~al.}(2004)\citenamefont {Keren},
  \citenamefont {Gardner}, \citenamefont {Ehlers}, \citenamefont {Fukaya},
  \citenamefont {Segal},\ and\ \citenamefont {Uemura}}]{Keren107204}%
  \BibitemOpen
  \bibfield  {author} {\bibinfo {author} {\bibfnamefont {A.}~\bibnamefont
  {Keren}}, \bibinfo {author} {\bibfnamefont {J.~S.}\ \bibnamefont {Gardner}},
  \bibinfo {author} {\bibfnamefont {G.}~\bibnamefont {Ehlers}}, \bibinfo
  {author} {\bibfnamefont {A.}~\bibnamefont {Fukaya}}, \bibinfo {author}
  {\bibfnamefont {E.}~\bibnamefont {Segal}},\ and\ \bibinfo {author}
  {\bibfnamefont {Y.~J.}\ \bibnamefont {Uemura}},\ }\bibfield  {title}
  {\bibinfo {title} {Dynamic properties of a diluted pyrochlore cooperative
  paramagnet ({Tb}$_p${Y}$_{1-p}$)$_2${Ti}$_2${O}$_7$},\ }\href
  {https://doi.org/10.1103/PhysRevLett.92.107204} {\bibfield  {journal}
  {\bibinfo  {journal} {Phys. Rev. Lett.}\ }\textbf {\bibinfo {volume} {92}},\
  \bibinfo {pages} {107204} (\bibinfo {year} {2004})}\BibitemShut {NoStop}%
\bibitem [{\citenamefont {Helton}\ \emph {et~al.}(2010)\citenamefont {Helton},
  \citenamefont {Matan}, \citenamefont {Shores}, \citenamefont {Nytko},
  \citenamefont {Bartlett}, \citenamefont {Qiu}, \citenamefont {Nocera},\ and\
  \citenamefont {Lee}}]{Helton147201}%
  \BibitemOpen
  \bibfield  {author} {\bibinfo {author} {\bibfnamefont {J.~S.}\ \bibnamefont
  {Helton}}, \bibinfo {author} {\bibfnamefont {K.}~\bibnamefont {Matan}},
  \bibinfo {author} {\bibfnamefont {M.~P.}\ \bibnamefont {Shores}}, \bibinfo
  {author} {\bibfnamefont {E.~A.}\ \bibnamefont {Nytko}}, \bibinfo {author}
  {\bibfnamefont {B.~M.}\ \bibnamefont {Bartlett}}, \bibinfo {author}
  {\bibfnamefont {Y.}~\bibnamefont {Qiu}}, \bibinfo {author} {\bibfnamefont
  {D.~G.}\ \bibnamefont {Nocera}},\ and\ \bibinfo {author} {\bibfnamefont
  {Y.~S.}\ \bibnamefont {Lee}},\ }\bibfield  {title} {\bibinfo {title} {Dynamic
  scaling in the susceptibility of the spin-$\frac{1}{2}$ kagome lattice
  antiferromagnet herbertsmithite},\ }\href
  {https://doi.org/10.1103/PhysRevLett.104.147201} {\bibfield  {journal}
  {\bibinfo  {journal} {Phys. Rev. Lett.}\ }\textbf {\bibinfo {volume} {104}},\
  \bibinfo {pages} {147201} (\bibinfo {year} {2010})}\BibitemShut {NoStop}%
\bibitem [{\citenamefont {Pourovskii}\ \emph {et~al.}(2026)\citenamefont
  {Pourovskii}, \citenamefont {Soares},\ and\ \citenamefont
  {Wietek}}]{abinitio}%
  \BibitemOpen
  \bibfield  {author} {\bibinfo {author} {\bibfnamefont {L.~V.}\ \bibnamefont
  {Pourovskii}}, \bibinfo {author} {\bibfnamefont {R.~D.}\ \bibnamefont
  {Soares}},\ and\ \bibinfo {author} {\bibfnamefont {A.}~\bibnamefont
  {Wietek}},\ }\bibfield  {title} {\bibinfo {title} {Ab initio spin
  hamiltonians and magnetism of ce and yb triangular-lattice compounds},\
  }\href {https://doi.org/10.1103/95x5-3fnn} {\bibfield  {journal} {\bibinfo
  {journal} {Phys. Rev. B}\ }\textbf {\bibinfo {volume} {113}},\ \bibinfo
  {pages} {L060401} (\bibinfo {year} {2026})}\BibitemShut {NoStop}%
\bibitem [{\citenamefont {Shimokawa}\ \emph {et~al.}(2025)\citenamefont
  {Shimokawa}, \citenamefont {Sabharwal},\ and\ \citenamefont
  {Shannon}}]{shimokawa2025}%
  \BibitemOpen
  \bibfield  {author} {\bibinfo {author} {\bibfnamefont {T.}~\bibnamefont
  {Shimokawa}}, \bibinfo {author} {\bibfnamefont {S.}~\bibnamefont
  {Sabharwal}},\ and\ \bibinfo {author} {\bibfnamefont {N.}~\bibnamefont
  {Shannon}},\ }\bibfield  {title} {\bibinfo {title} {Can
  ex\-per\-i\-men\-tal\-ly\--accessible measures of entanglement distinguish
  quantum spin liquids from disorder-driven ``random singlet'' phases?},\
  }\Eprint {https://arxiv.org/abs/2505.11874} {arXiv:2505.11874
  [cond-mat.str-el]}  (\bibinfo {year} {2025})\BibitemShut {NoStop}%
\bibitem [{\citenamefont {Wu}\ \emph {et~al.}(2019)\citenamefont {Wu},
  \citenamefont {Gong},\ and\ \citenamefont {Sheng}}]{wu2019randomness}%
  \BibitemOpen
  \bibfield  {author} {\bibinfo {author} {\bibfnamefont {H.-Q.}\ \bibnamefont
  {Wu}}, \bibinfo {author} {\bibfnamefont {S.-S.}\ \bibnamefont {Gong}},\ and\
  \bibinfo {author} {\bibfnamefont {D.~N.}\ \bibnamefont {Sheng}},\ }\bibfield
  {title} {\bibinfo {title} {Randomness-in\-duced spin-liquid-like phase in the
  spin-$\frac{1}{2}$ ${J}_{1}$-${J}_{2}$ triangular {H}eisenberg model},\
  }\href {https://doi.org/10.1103/PhysRevB.99.085141} {\bibfield  {journal}
  {\bibinfo  {journal} {Phys. Rev. B}\ }\textbf {\bibinfo {volume} {99}},\
  \bibinfo {pages} {085141} (\bibinfo {year} {2019})}\BibitemShut {NoStop}%
\bibitem [{\citenamefont {Chatterjee}\ \emph {et~al.}(2023)\citenamefont
  {Chatterjee}, \citenamefont {Puphal}, \citenamefont {Barthélemy},
  \citenamefont {Willwater}, \citenamefont {Süllow}, \citenamefont {Baines},
  \citenamefont {Petit}, \citenamefont {Ressouche}, \citenamefont {Ollivier},
  \citenamefont {Zoch}, \citenamefont {Krellner}, \citenamefont {Parzer},
  \citenamefont {Riss}, \citenamefont {Garmroudi}, \citenamefont {Pustogow},
  \citenamefont {Mendels}, \citenamefont {Kermarrec},\ and\ \citenamefont
  {Bert}}]{Chatterjee2023}%
  \BibitemOpen
  \bibfield  {author} {\bibinfo {author} {\bibfnamefont {D.}~\bibnamefont
  {Chatterjee}}, \bibinfo {author} {\bibfnamefont {P.}~\bibnamefont {Puphal}},
  \bibinfo {author} {\bibfnamefont {Q.}~\bibnamefont {Barthélemy}}, \bibinfo
  {author} {\bibfnamefont {J.}~\bibnamefont {Willwater}}, \bibinfo {author}
  {\bibfnamefont {S.}~\bibnamefont {Süllow}}, \bibinfo {author} {\bibfnamefont
  {C.}~\bibnamefont {Baines}}, \bibinfo {author} {\bibfnamefont
  {S.}~\bibnamefont {Petit}}, \bibinfo {author} {\bibfnamefont
  {E.}~\bibnamefont {Ressouche}}, \bibinfo {author} {\bibfnamefont
  {J.}~\bibnamefont {Ollivier}}, \bibinfo {author} {\bibfnamefont {K.~M.}\
  \bibnamefont {Zoch}}, \bibinfo {author} {\bibfnamefont {C.}~\bibnamefont
  {Krellner}}, \bibinfo {author} {\bibfnamefont {M.}~\bibnamefont {Parzer}},
  \bibinfo {author} {\bibfnamefont {A.}~\bibnamefont {Riss}}, \bibinfo {author}
  {\bibfnamefont {F.}~\bibnamefont {Garmroudi}}, \bibinfo {author}
  {\bibfnamefont {A.}~\bibnamefont {Pustogow}}, \bibinfo {author}
  {\bibfnamefont {P.}~\bibnamefont {Mendels}}, \bibinfo {author} {\bibfnamefont
  {E.}~\bibnamefont {Kermarrec}},\ and\ \bibinfo {author} {\bibfnamefont
  {F.}~\bibnamefont {Bert}},\ }\bibfield  {title} {\bibinfo {title} {{From spin
  liquid to magnetic ordering in the anisotropic kagome Y-kapellasite
  Y$_3$Cu$_9$(OH)$_{19}$Cl$_8$ : A single-crystal study}},\ }\href
  {https://doi.org/10.1103/physrevb.107.125156} {\bibfield  {journal} {\bibinfo
   {journal} {Physical Review B}\ }\textbf {\bibinfo {volume} {107}},\ \bibinfo
  {pages} {125156} (\bibinfo {year} {2023})}\BibitemShut {NoStop}%
\bibitem [{\citenamefont {Zeng}\ \emph {et~al.}(2022)\citenamefont {Zeng},
  \citenamefont {Ma}, \citenamefont {Wu}, \citenamefont {Li}, \citenamefont
  {Tao}, \citenamefont {Lu}, \citenamefont {Chen}, \citenamefont {Mi},
  \citenamefont {Song}, \citenamefont {Cao}, \citenamefont {Che}, \citenamefont
  {Li}, \citenamefont {Li}, \citenamefont {Luo}, \citenamefont {Meng},\ and\
  \citenamefont {Li}}]{PhysRevB.105.L121109}%
  \BibitemOpen
  \bibfield  {author} {\bibinfo {author} {\bibfnamefont {Z.}~\bibnamefont
  {Zeng}}, \bibinfo {author} {\bibfnamefont {X.}~\bibnamefont {Ma}}, \bibinfo
  {author} {\bibfnamefont {S.}~\bibnamefont {Wu}}, \bibinfo {author}
  {\bibfnamefont {H.-F.}\ \bibnamefont {Li}}, \bibinfo {author} {\bibfnamefont
  {Z.}~\bibnamefont {Tao}}, \bibinfo {author} {\bibfnamefont {X.}~\bibnamefont
  {Lu}}, \bibinfo {author} {\bibfnamefont {X.}~\bibnamefont {Chen}}, \bibinfo
  {author} {\bibfnamefont {J.-X.}\ \bibnamefont {Mi}}, \bibinfo {author}
  {\bibfnamefont {S.-J.}\ \bibnamefont {Song}}, \bibinfo {author}
  {\bibfnamefont {G.-H.}\ \bibnamefont {Cao}}, \bibinfo {author} {\bibfnamefont
  {G.}~\bibnamefont {Che}}, \bibinfo {author} {\bibfnamefont {K.}~\bibnamefont
  {Li}}, \bibinfo {author} {\bibfnamefont {G.}~\bibnamefont {Li}}, \bibinfo
  {author} {\bibfnamefont {H.}~\bibnamefont {Luo}}, \bibinfo {author}
  {\bibfnamefont {Z.~Y.}\ \bibnamefont {Meng}},\ and\ \bibinfo {author}
  {\bibfnamefont {S.}~\bibnamefont {Li}},\ }\bibfield  {title} {\bibinfo
  {title} {Possible {D}irac quantum spin liquid in the kagome quantum
  antiferromagnet {YCu}$_3$\-({OH})$_6$\-{Br}$_2$\-[{Br}$_x$({OH})$_{1-x}$]},\
  }\href {https://doi.org/10.1103/PhysRevB.105.L121109} {\bibfield  {journal}
  {\bibinfo  {journal} {Phys. Rev. B}\ }\textbf {\bibinfo {volume} {105}},\
  \bibinfo {pages} {L121109} (\bibinfo {year} {2022})}\BibitemShut {NoStop}%
\bibitem [{\citenamefont {Wysin}(2012)}]{wysin2012demag}%
  \BibitemOpen
  \bibfield  {author} {\bibinfo {author} {\bibfnamefont {G.~M.}\ \bibnamefont
  {Wysin}},\ }\bibfield  {title} {\bibinfo {title} {Demagnetization fields}}
  (\bibinfo {year} {2012}),\ \bibinfo {note} {{\it unpublished},
  \url{http://www.phys.ksu.edu/personal/wysin}}\BibitemShut {NoStop}%
\end{thebibliography}

\end{document}